\documentclass[aps,prd,showpacs,floatfix,nofootinbib,twocolumn,10pt]{revtex4}

\usepackage{color,amsmath,amssymb,graphicx,latexsym,subfigure}
\usepackage{threeparttable,txfonts}

\newcommand{\sv}{\ensuremath{\langle\sigma v\rangle}}

\begin{document}

\title{Interpretations of the DAMPE electron data}

\author{Qiang Yuan$^{a,b}$}
\author{Lei Feng$^a$\footnote{Co-first author.}}
\author{Peng-Fei Yin$^c$}
\author{Yi-Zhong Fan$^{a,b}$\footnote{Corresponding author: yzfan@pmo.ac.cn}}
\author{Xiao-Jun Bi$^c$\footnote{Corresponding author: bixj@ihep.ac.cn}}
\author{Ming-Yang Cui$^{a,e}$}
\author{Tie-Kuang Dong$^a$}
\author{Yi-Qing Guo$^c$}
\author{Kun Fang$^c$}
\author{Hong-Bo Hu$^c$}
\author{Xiaoyuan Huang$^d$}
\author{Shi-Jun Lei$^a$}
\author{Xiang Li$^a$}
\author{Su-Jie Lin$^c$}
\author{Hao Liu$^a$}
\author{Peng-Xiong Ma$^{a,b}$}
\author{Wen-Xi Peng$^{c}$}
\author{Rui Qiao$^{c}$}
\author{Zhao-Qiang Shen$^{a,f}$}
\author{Meng Su$^{a,g}$}
\author{Yi-Feng Wei$^h$}
\author{Zun-Lei Xu$^{a,f}$}
\author{Chuan Yue$^{a,f}$}
\author{Jing-Jing Zang$^a$}
\author{Cun Zhang$^{a,e}$}
\author{Xinmin Zhang$^i$}
\author{Ya-Peng Zhang$^j$}
\author{Yong-Jie Zhang$^j$}
\author{Yun-Long Zhang$^h$}

\affiliation{
$^a$Key Laboratory of Dark Matter and Space Astronomy, Purple Mountain
Observatory, Chinese Academy of Sciences, Nanjing 210008, China \\
$^b$School of Astronomy and Space Science, University of Science and
Technology of China, Hefei 230026, China\\
$^c$Key Laboratory of Particle Astrophysics, Institute of High Energy
Physics, Chinese Academy of Sciences, Beijing 100049, China\\
$^d$Niels Bohr Institute, University of Copenhagen, Blegdamsvej 17,
2100 Copenhagen {\O}, Denmark\\
$^e$School of Physics, Nanjing University, Nanjing, 210092, China\\
$^f$University of Chinese Academy of Sciences, 19 Yuquan Road, Beijing
100049, China\\
$^g$Department of Physics and Laboratory for Space Research,
University of Hong Kong, Pokfulam Road, Hong Kong\\
$^h$State Key Laboratory of Particle Detection and Electronics, University
of Science and Technology of China, Hefei 230026, China\\
$^i$Theoretical Physics Division, Institute of High Energy Physics,
Chinese Academy of Sciences, Beijing 100049, China\\
$^j$Institute of Modern Physics, Chinese Academy of Sciences, Lanzhou
730000, China
}

\begin{abstract}

The DArk Matter Particle Explorer (DAMPE), a high energy cosmic ray and
$\gamma$-ray detector in space, has recently reported the new measurement
of the total electron plus positron flux between 25 GeV and 4.6 TeV. 
A spectral softening at $\sim0.9$ TeV and a tentative peak at $\sim1.4$ 
TeV have been reported. We study the physical implications of the DAMPE 
data in this work. The presence of the spectral break significantly tightens 
the constraints on the model parameters to explain the electron/positron
excesses. The spectral softening can either be explained by the maximum
acceleration limits of electrons by astrophysical sources, or a breakdown 
of the common assumption of continuous distribution of electron sources at
TeV energies in space and time. The tentive peak at $\sim1.4$ TeV implies 
local sources of electrons/positrons with quasi-monochromatic injection 
spectrum. We find that the cold, ultra-relativistic $e^+e^-$ winds from 
pulsars may give rise to such a structure. The pulsar is requird to be 
middle-aged, relatively slowly-rotated, mildly magnetized, and isolated in 
a density cavity. The annihilation of DM particles ($m_{\chi}\sim1.5$ TeV) 
into $e^+e^-$ pairs in a nearby clump or an over-density region may also 
explain the data. In the DM scenario, the inferred clump mass (or density
enhancement) is about $10^7-10^8$ M$_\odot$ (or $17-35$ times of the
canonical local density) assuming a thermal production cross section,
which is relatively extreme compared with the expectation from numerical 
simulations. A moderate enhancement of the annihilation cross section 
via, e.g., the Sommerfeld mechanism or non-thermal production,
is thus needed.

\end{abstract}

\date{\today}


\maketitle

\section{Introduction}

High energy electrons and positrons are very important probe of nearby
cosmic ray (CR) sources (e.g., pulsars \cite{1970ApJ...162L.181S,
1987ICRC....2...92H,1995A&A...294L..41A,1996ApJ...459L..83C,
2001A&A...368.1063Z}) as well as the particle dark matter (DM; e.g., 
\cite{2000RPPh...63..793B,2005PhR...405..279B,2009NJPh...11j5006B}). 
Recent discoveries of the excesses of positrons \cite{2009Natur.458..607A,
2012PhRvL.108a1103A,2013PhRvL.110n1102A,2014PhRvL.113l1101A} and electrons
\cite{2008Natur.456..362C,2008PhRvL.101z1104A,2009PhRvL.102r1101A,
2014PhRvL.113l1102A,2014PhRvL.113v1102A} stimulated quite a number of
works to discuss their possible origin, either astrophysical sources (see
e.g., \cite{2010IJMPD..19.2011F,2012APh....39....2S}) or the DM
annihilation or decay (e.g., \cite{2008PhRvD..78j3520B,2009JCAP...08..017I,
2012Prama..79.1021C,2013FrPhy...8..794B,2016ConPh..57..496G}). 
It has been shown that pulsars may explain the data well 
\cite{2009JCAP...01..025H,2009PhRvL.103e1101Y,2012CEJPh..10....1P}. 
If the DM annihilation or decay is employed to explain the data, then 
only in a few cases with flat density profile and/or leptonic 
annihilation/decay channel the model can be consistent with $\gamma$-ray 
and antiproton observations \cite{2009NuPhB.813....1C,2009PhRvD..79b3512Y,
2009JCAP...03..009B,2009PhRvD..80b3007Z,2010JCAP...03..014P}.

TeV electrons can only travel by a small distance ($\sim$kpc) in the Milky
Way due to strong radiative cooling via synchrotron and inverse Compton
scattering (ICS) processes. Therefore, the electron spectrum up to TeV
energies is expected to reveal directly the origin and transportation of
electrons in the local Galaxy. In particular, the continuous source
distribution of electrons (both in space and time) is expected to be
violated at such high energies, and the local, perhaps fresh, sources
play a significant role in regulating the TeV spectrum of electrons
\cite{1995A&A...294L..41A}. The inferred primary electron spectral
hardening from the AMS-02 data \cite{2013PhRvL.110n1102A,
2014PhRvL.113l1101A,2014PhRvL.113l1102A,2014PhRvL.113v1102A} may be an
indication of the breakdown of continuous source distribution
\cite{2014PhLB..728..250F,2015APh....60....1Y,
2013PhRvD..88b3013C,2013PhLB..727....1Y,2015PhLB..749..267L}.
The measurement of the electron spectrum to even higher energies with
improved precision is thus crucial to further test this continuous source
assumption, probe the nearby Galactic environment and/or even identify TeV
electron sources.

The DArk Matter Particle Explorer (DAMPE; \cite{ChangJin:550,
2017APh....95....6C}) has recently measured the total electron plus 
positron fluxes up to 4.6 TeV with unprecedentedly high quality
\cite{DAMPE-2017CRE}. The energy resolution of the DAMPE is better than 
$1.5\%$ at TeV energies, and the hadron rejection power is about $10^5$ 
\cite{2017APh....95....6C}. Such excellent performance enables DAMPE to 
reveal (fine) structures of the $e^++e^-$ fluxes. The DAMPE data display 
a spectral softening at $\sim0.9$ TeV and a tentative peak at $\sim1.4$ 
TeV \cite{DAMPE-2017CRE}. The spectral softening may be due to the 
breakdown of the conventional assumption of continuous source 
distribution or the maximum acceleration limits of electron sources. 
The peak structure at $\sim1.4$ TeV, although its significance is not 
high, is more challenging to be understood. The energy density is 
estimated to be about $1.2\times10^{-18}$ erg cm$^{-3}$. 
To produce such a structure, nearly monochromatic injection of electrons 
is required. Furthermore, the source should be young and close enough to 
the Earth that cooling is not significant to modify the injection 
spectrum\footnote{Strictly speaking, this depends on instantaneous or 
continuous injection of the source. For instantaneous injection, the 
cooling might not broaden the injection spectrum.}. The cooling time of 
1 TeV electrons in the local interstellar environment is about 
$3\times10^5$ yr, which corresponds to a diffusion length of $\sim1$ kpc. 
Therefore, the DAMPE peak should be predominantly originated from 
late-time injection of nearby sources.

In this work we study the implications of the DAMPE data on our
understanding of high energy CR electron sources. Either astrophysical
sources (e.g., pulsars) or exotic sources (e.g., the annihilation or
deday of DM) will be discussed. We infer the properties and parameters
of the sources from fitting to the data, and employ other kinds of
observations such as $\gamma$-rays to further test or constrain the models.

The discussion consists of two parts: the updated constraints on models 
to explain the electron/positron excesses in light of AMS-02 and DAMPE 
data, and the interpretations of the potential spectral feature
by DAMPE. The paper is outlined as follows. In Sec. II we briefly
introduce the propagation of electrons in the Milky Way. In Sec. III
we study the implications on the background and extra sources of CR
electrons/positrons from the wide-band AMS-02 and DAMPE data. The models
to explain the peak of DAMPE are investigated in Sec. IV. We discuss 
the anisotropy which may distinguish different models in Sec. V, 
and conclude in Sec. VI.

\section{Propagation of CR electrons}

Charged CRs propagate diffusively in the turbulent magnetic field of
the Milky Way. For electrons, a distinct property of the propagation is
the radiative cooling, which is especially important at high energies.
The electron cooling rate in the local environment can be approximated
as \cite{1995PhRvD..52.3265A}
\begin{equation}
-{\rm d}E/{\rm d}t\equiv b(E)=b_0+b_1E_{\rm GeV}+b_2E_{\rm GeV}^2,
\label{be}
\end{equation}
where $E_{\rm GeV}\equiv E/{\rm GeV}$. The first term in the right hand
side, with $b_0\approx3\times10^{-16}$ GeV s$^{-1}$, represents the
ionization loss rate in a neutral gas with a density of 1 cm$^{-3}$,
the middle term is the bremsstrahlung loss in the same neutral gas, with
$b_1\approx10^{-15}$ GeV s$^{-1}$, and the last term is the synchrotron
and ICS losses with $b_2\approx1.0\times10^{-16}$ GeV s$^{-1}$ for a sum
energy density of 1 eV cm$^{-3}$ for both the magnetic field and
insterstellar radiation field. The cooling time of electrons is defined
as $\tau(E)\equiv E/b(E)$, and the effective propagation length of an
electron within its cooling time can be estimated as
\begin{equation}
\lambda(E)=2\left(\int_E^{\infty}\frac{D(E')}{b(E')}dE'\right)^{1/2},
\end{equation}
where $D(E)$ is the spatial diffusion coefficient.
The strong cooling makes high energy electrons originate locally from
the source. For typical diffusion parameters (see below), we find that
the cooling time (propagation length) varies from 10 Myr (10 kpc) for
GeV electrons to $<$Myr ($\sim$kpc) for TeV energies. For the energies
we are mostly interested in, i.e., from hundreds of GeV to TeV, the
electrons should dominately come from a volume of $\sim$kpc$^3$.

Due to the local origin of high energy electrons, the propagation equation
can be solved analytically, assuming a spherically symmetric geometry with
infinite boundary conditions (see \cite{1995PhRvD..52.3265A} for details).
For low energy electrons, however, the spherically symmetric solution
may not be proper any longer. In this work, we employ the numerical
tool {\tt GALPROP}\footnote{http://sourceforge.net/projects/galprop/}
\cite{1998ApJ...509..212S,1998ApJ...493..694M} to solve the propagation
of such low energy, ``background'' electrons, defined as contribution
from a population of sources continuously distributed in the Galaxy.
The spherical Green's function will be used when isolate, nearby source(s)
are discussed. The benchmark propagation parameters are
\cite{2012ApJ...761...91A}: the diffusion coefficient $D(R)=\beta
D_0(R/{\rm GV})^{\delta}$ with $D_0=3.3\times 10^{28}$ cm$^2$ s$^{-1}$
and $\delta=1/3$, the half height of the propagation cylinder $z_h=4$ kpc,
and the Alfvenic speed of the medium $v_A=33.5$ km s$^{-1}$.



\section{Implications on the background and extra sources of
electrons/positrons from DAMPE and AMS-02}

The DAMPE measurements extend the total $e^++e^-$ spectrum to multi-TeV
energies with high precision, and hence more stringent constraints
on the model parameters, for either the background or the extra sources,
are expected. In this section we study the implications of the wide
band behaviors of the electron/positron spectra from DAMPE and AMS-02.

\subsection{Fitting method}

We adopt the {\tt CosRayMC} tool, which embeds the CR propagation
tool {\tt GALPROP} into a Markov Chain Monte Carlo (MCMC) sampler
\cite{2002PhRvD..66j3511L}, to fit the data. More detailed description
of the code can be found in Refs. \cite{2010PhRvD..81b3516L,
2012PhRvD..85d3507L}. Compared with the original version of the
{\tt CosRayMC}, we further employ a ``Green's function'' method
to enable fast computation of the propagation given the spatial
distribution of the sources \cite{2017CoPhC.213..252H}.

We simply outline the model configuration here.
\begin{itemize}
\item The background electrons, assumed to be accelerated simultaneously
with the primary nuclei, are injected into the Galaxy with a three-piece
broken power-law spectrum. The first break at a few GeV is to fit the
low energy data, and the second one at $O(100)$ GeV (a spectral hardening,
probably due to nearby sources \cite{2013A&A...555A..48B} or non-linear
particle acceleration \cite{2013ApJ...763...47P}) is inferred via a global
fittting to the positron and electron data
\cite{2014PhLB..728..250F,2015APh....60....1Y,2013PhLB..727....1Y}.
Furthermore, we add an exponential cutoff of the background electron
spectrum at $\sim$TeV, to reproduce the drop observed by DAMPE. The
spatial distribution of background electrons is assumed to follow the
supernova remnant (SNR) distribution with parameters adjusted to match
the diffuse $\gamma$-ray data \cite{2011ApJ...729..106T}.

\item The background positrons are assumed to be produced by the inelastic
interactions between the primary nuclei and the interstellar medium (ISM)
during the propagation. We assume a free re-normalization parameter of
the background positrons, which describes possible uncertainties when
predicting the positron fluxes, from e.g., the determination of the
propagation parameters, the hadronic interaction cross section, and/or
the fluctuation of positrons in the Milky Way.

\item We assume two kinds of extra sources of electrons and positrons.
The astrophysical sources are represented by a population of pulsars,
whose spatial distribution is similar with the background CR sources
as described above. The injection spectrum of electrons/positrons from
pulsars is parameterized as an exponential cutoff power-law form. The
annihilation or decay of DM particles will also be discussed. The DM
density profile is assumed to be an Navarro-Frenk-White (NFW;
\cite{1997ApJ...490..493N}) distribution, $\rho(r)=\rho_s\left[(r/r_s)
(1+r/r_s)^2\right]^{-1}$, with a scale radius of $r_s=20$ kpc and a
local density of $\rho_0=\rho(r_{\odot})=0.4$ GeV cm$^{-3}$.
The production spectrum of positrons is calculated according to the
tables given in Ref. \cite{2011JCAP...03..051C}.

\item After entering the heliosphere, the low energy particles will
be modulated by solar magnetic field. The force-field approximation of
the solar modulation is assumed \cite{1968ApJ...154.1011G}. The modulation
potential $\Phi$ is treated as a free parameter in the fittings.

\end{itemize}
The data used in the fittings include the positron fraction and total
$e^++e^-$ spectrum measured by AMS-02 \cite{2014PhRvL.113l1101A,
2014PhRvL.113v1102A}, the total $e^++e^-$ spectrum by HESS
\cite{2008PhRvL.101z1104A}, and/or the DAMPE data. The positron fraction
by AMS-02 is fitted simultaneously with other data sets, because the flux
ratio is expected to have lower systematics. Note that new measurements
of the total $e^++e^-$ spectrum up to 2 TeV by Fermi-LAT
\cite{2017PhRvD..95h2007A}, which are in general agreement with that
of DAMPE, have not been included in the fittings. Furthermore, the
AMS-02 data below 1 GeV are eliminated, because they are strongly
affected by the solar modulation and may not be well reproduced by the
force field model. When the DAMPE data are included, the AMS-02 data
above $25$ GeV whose energy coverage overlaps with that of DAMPE but
with slightly different absolute fluxes, are excluded. We further exclude
the one DAMPE data point around $\sim1.4$ TeV in the fittings, which
reveals spiky structure and will be discussed separately.

\subsection{Astrophysical sources}

Astrophysical sources such as pulsars \cite{1970ApJ...162L.181S,
1987ICRC....2...92H,1996ApJ...459L..83C,2001A&A...368.1063Z,
2009JCAP...01..025H,2009PhRvL.103e1101Y,2009PhRvD..80f3005M,
2009APh....32..140G,2010ApJ...710..958K,2012CEJPh..10....1P,
2013ApJ...772...18L,2013PhRvD..88b3001Y,2016EPJC...76..229F,
2017ApJ...836..172F}, pulsar wind nebulae (PWNe;
\cite{2011APh....35..211L,2014JCAP...04..006D,2015A&A...575A..67B,
2015JHEAp...8...27D,2016JCAP...05..031D}), and/or SNRs
\cite{2009PhRvL.103e1104B,2009PhRvL.103k1302S,2009ApJ...700L.170H,
2009PhRvD..80f3003F,2010A&A...524A..51D,2016PTEP.2016b1E01K}) have
been widely discussed as origins of high energy electrons and positrons.
We take the pulsar scenario as an illustration for the dicsussion here.
The methodology is, however, applicable for other kinds of sources.

\begin{figure}[!htb]
\includegraphics[width=0.48\textwidth]{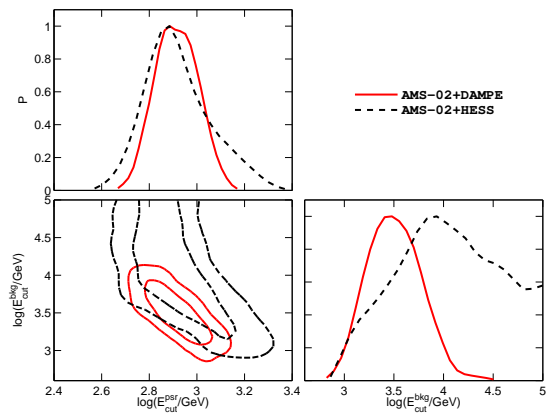}
\caption{Constraints on the cutoff energies of the background electrons
and electrons/positrons from pulsars. The bottom-left panel shows the
68\% and 95\% confidence regions of the two parameters, and the diagonal
panels show the one-dimensional probability distributions of them.
Black dashed lines and contours are for the fitting to the AMS-02 and
HESS data, while red solid ones are for the AMS-02 and DAMPE data.
\label{fig:psr_ecut}}
\end{figure}

Fig. \ref{fig:psr_ecut} shows the constraints on two parameters,
the cutoff of the background electron
spectrum $E_{\rm cut}^{\rm bkg}$, and the cutoff of the pulsar
injection spectrum, $E_{\rm cut}^{\rm psr}$, through fitting to the
AMS-02 + HESS data and AMS-02 + DAMPE data, respectively. Compared with
the fitting to the AMS-02 + HESS data, we find that the high energy
behaviors of both the background electrons and the pulsar component
can be constrained much better after including the DAMPE data.
Specifically, we get $E_{\rm cut}^{\rm bkg}=3.2^{+2.7}_{-1.5}$ TeV and
$E_{\rm cut}^{\rm psr}=0.82^{+0.19}_{-0.15}$ TeV, for the fitting to
the AMS-02 + DAMPE data. The cutoff of the background electrons can
not be effectively constrained for the fitting to the AMS-02 + HESS
data, primarily due to the large systematic uncertainties of the HESS
data \cite{2008PhRvL.101z1104A}.

\begin{figure*}[!htb]
\includegraphics[width=0.45\textwidth]{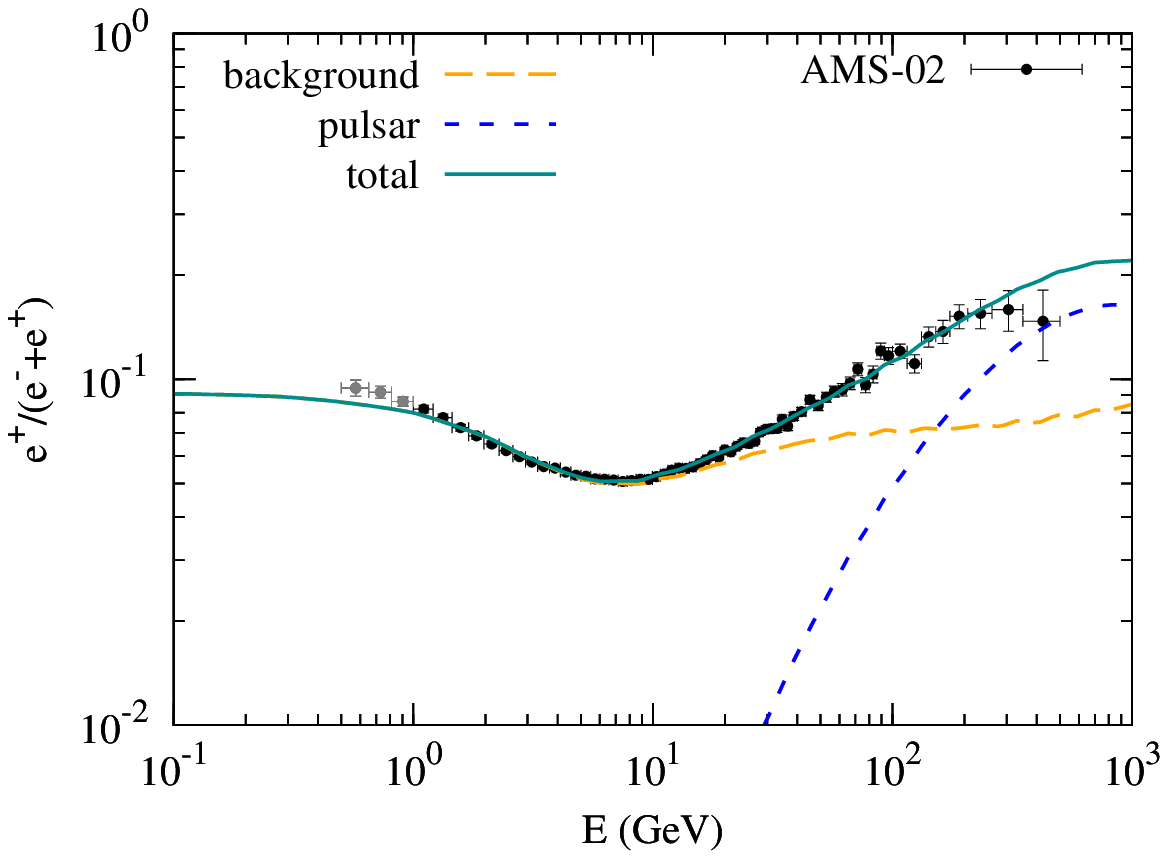}
\includegraphics[width=0.45\textwidth]{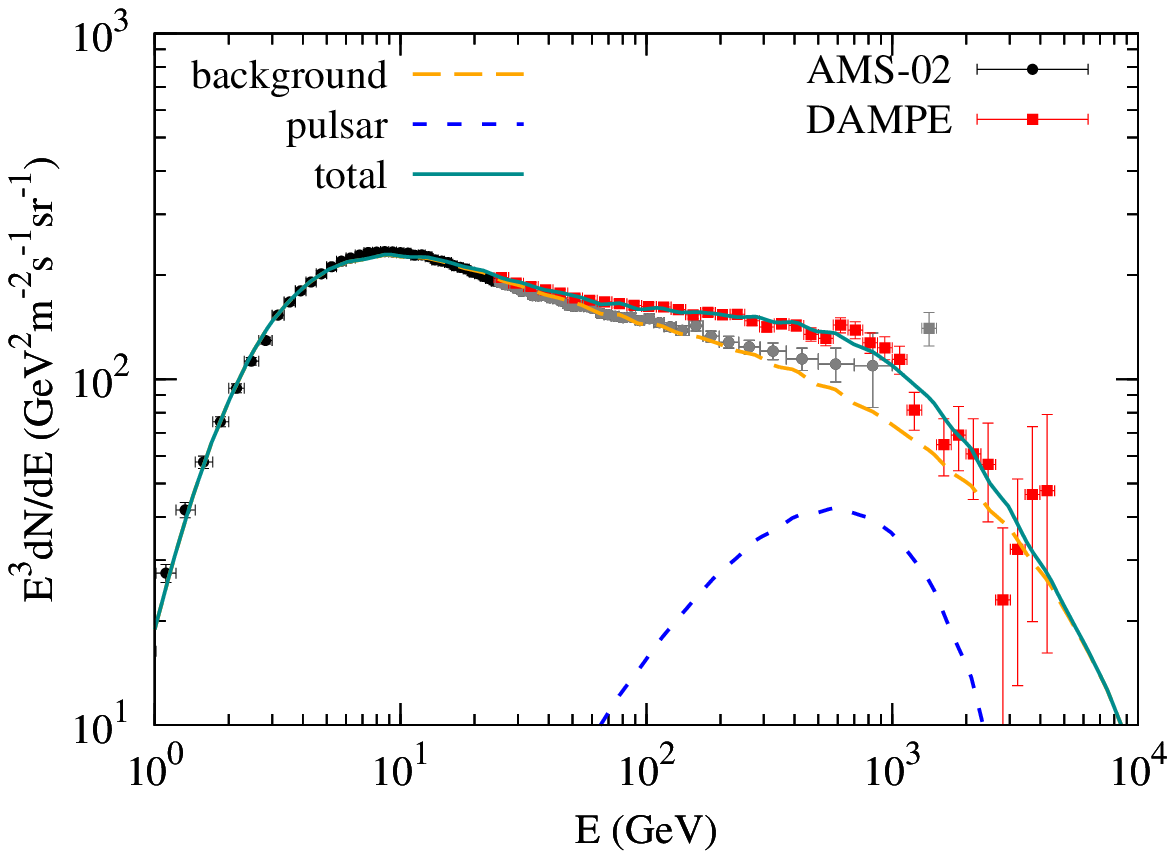}
\caption{The positron fraction (left) and total $e^++e^-$ fluxes (right)
for the pulsar model that best-fits the AMS-02 and DAMPE data. The data
points in gray are not included in the fitting.
\label{fig:psr}}
\end{figure*}

Fig. \ref{fig:psr} shows the best-fitting results of the positron fraction
(left) and total $e^++e^-$ fluxes (right) for the pulsar model, compared
with the data. We find that the model is well consistent with the data.

\subsection{DM annihilation or decay}

\begin{figure}[!htb]
\includegraphics[width=0.48\textwidth]{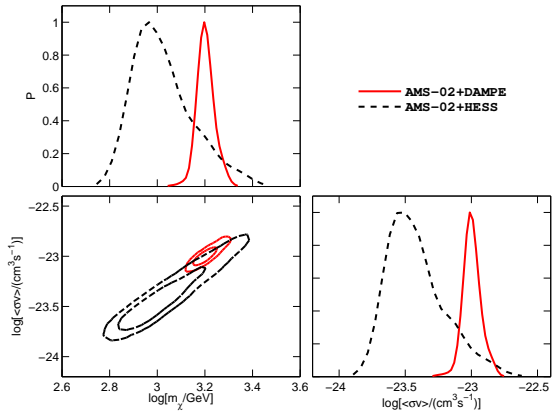}
\caption{Same as Fig. \ref{fig:psr_ecut} but for the constraints on the
DM mass and annihilation cross section, assuming $\mu^+\mu^-$ channel.
\label{fig:anni_mu_ecut}}
\end{figure}

The pair annihilation or decay of DM particles is widely postulated to
be source of CR electrons/positrons \cite{2000RPPh...63..793B,
2005PhR...405..279B,2009NJPh...11j5006B}. The DAMPE data are also expected
to improve the constraints on the DM model parameters effectively.
Fig. \ref{fig:anni_mu_ecut} shows the constraints on the mass and
cross section for the annihilating DM model, assuming $\mu^+\mu^-$
annihilation channel, for fittings to AMS-02 + HESS data and AMS-02 +
DAMPE data, respectively. We find that the DM parameters are much more
tightly constrained in comparison with the pre-DAMPE data.

\begin{table}[!htb]
\caption {Best-fit $\chi^2$ values of the DM models through fitting to
the AMS-02 + DAMPE data. The number of d.o.f. is 125.}
\begin{tabular}{ccccccc}
\hline \hline
channel & $e^+e^-$ & $\mu^+\mu^-$ & $\tau^+\tau^-$ & $4e$ & $4\mu$ & $4\tau$ \\
\hline
annihilation & 214.3 & 139.7 & 135.8 & 147.4 & 130.9 & 133.8 \\
decay & 215.6 & 140.3 & 128.6 & 160.4 & 126.9 & 126.2 \\
\hline \hline
\end{tabular}
\label{table:chi2}
\end{table}

\begin{figure*}[!htb]
\includegraphics[width=0.45\textwidth]{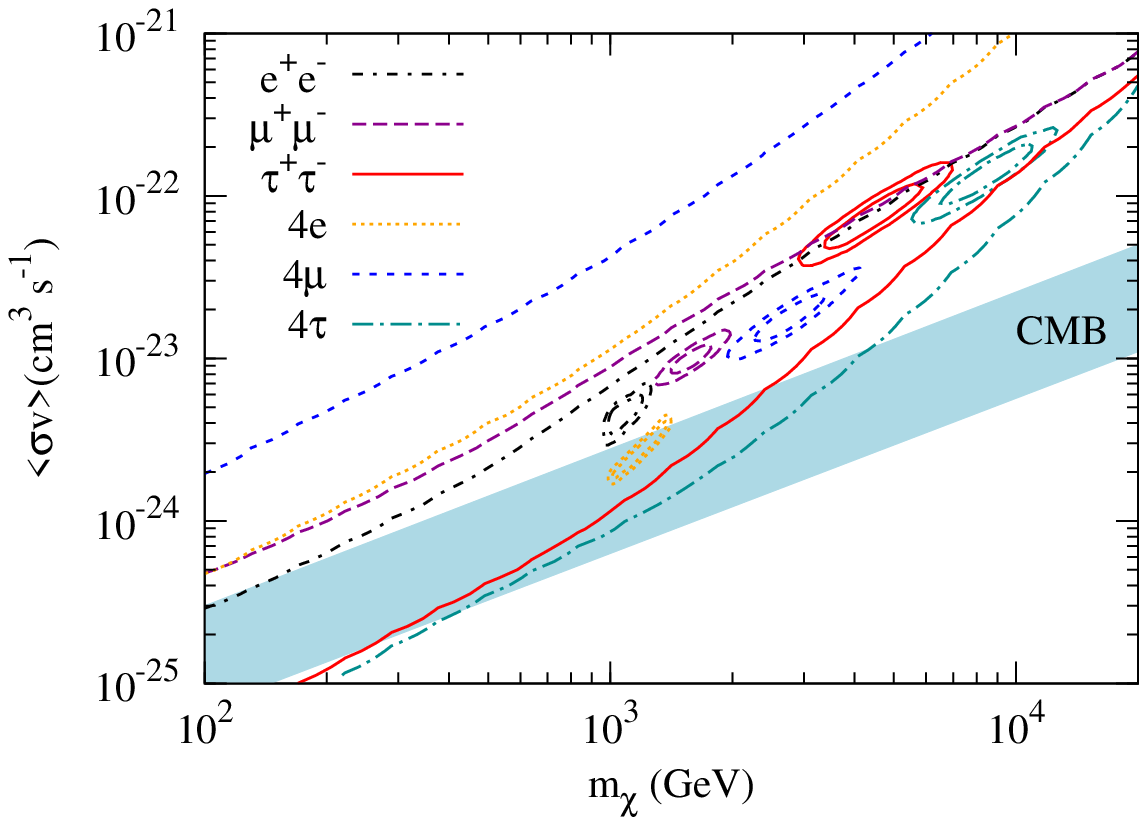}
\includegraphics[width=0.45\textwidth]{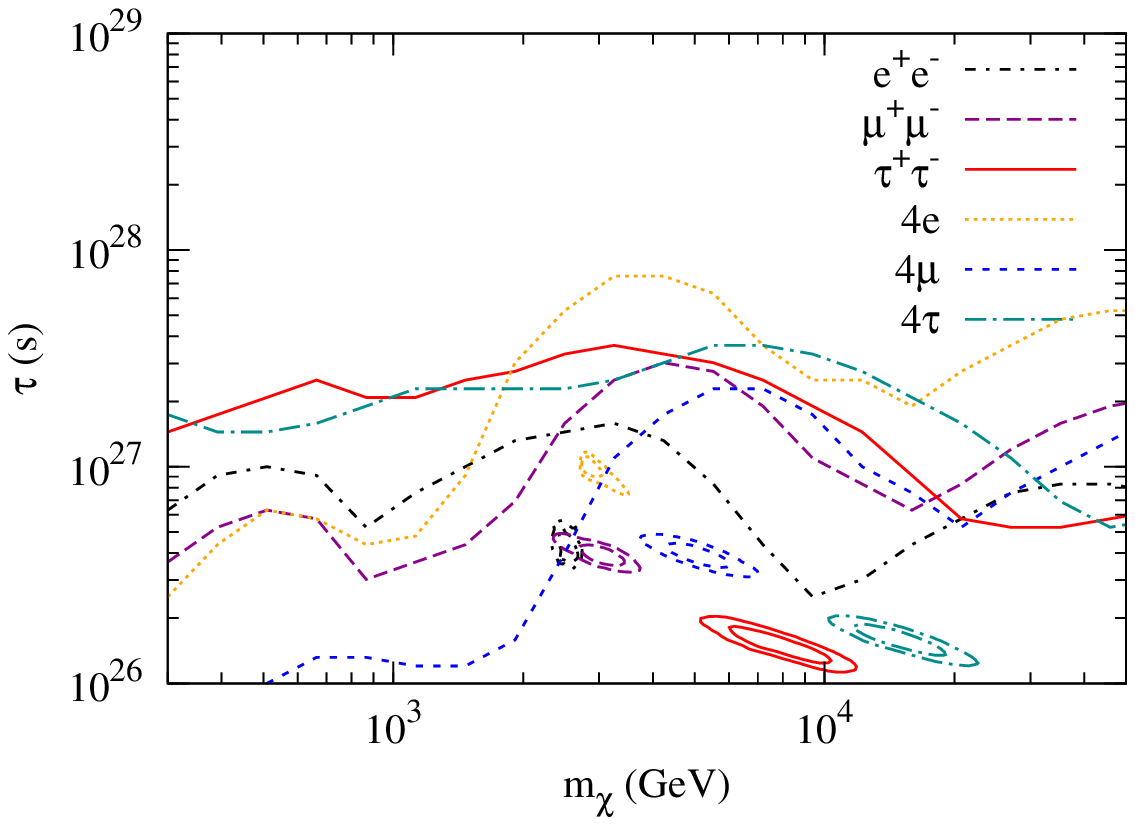}
\caption{Favored parameter regions on the $m_{\chi}-\sv$ (for annihilation;
left) and $m_{\chi}-\tau$ (for decay; right) plane to explain the CR
electron/positron data, for channels $e^+e^-$, $\mu^+\mu^-$, $\tau^+\tau^-$,
$4e$, $4\mu$, and $4\tau$, respectively. The shaded region in the left
panel shows the constraints from the Planck observations of CMB anisotropies,
assuming an energy deposition efficiency, $f_{\rm eff}$, of $\sim0.7$
(for $e^+e^-$ channel) to $\sim0.15$ (for $\tau^+\tau^-$ channel)
\cite{2016A&A...594A..13P}.
\label{fig:sv-tau}}
\end{figure*}

We also consider other leptonic channels of DM annihilation or decay
to account for the AMS-02 and DAMPE data. Fig. \ref{fig:sv-tau} presents
the fitting $68\%$ and $95\%$ parameter regions of $m_{\chi}$ and $\sv$
(for annihilating DM) or $\tau$ (for decaying DM), for $e^+e^-$,
$\mu^+\mu^-$, $\tau^+\tau^-$, $4e$, $4\mu$, and $4\tau$ channels.
For the four-lepton cases, DM particles are assumed to first annihilate
or decay into a pair of intermediate, light particles $\phi$ ($m_{\phi}
<$ a few GeV), each of which decays quickly into a pair of leptons.
All of these leptonic channels can give reasonably good fittings to the
data. The reduced $\chi^2$ values of the fittings are about $1.0-1.3$
except for the $e^+e^-$ channel, for a number of degree-of-freedom
(d.o.f.) of 125, as tabulated in Table \ref{table:chi2}. The annihilation
or decay into $e^+e^-$ fits the data poorly, because the electron/positron
spectrum is very hard that the AMS-02 positron fraction data requires a
low DM mass, which is difficult to explain the DAMPE data above $\sim$TeV.

We compare the parameters derived in this work with that given in
Ref. \cite{2015JCAP...03..033Y} in which similar fittings to the AMS-02
and Fermi-LAT data were done\footnote{Note that in Ref.
\cite{2015JCAP...03..033Y} the local density of DM was assumed to be 0.3
GeV cm$^{-3}$, hence the cross section is higher than that of this work
for the same DM mass.}. We find that the mass of DM particles required
to account for the DAMPE data is larger by a factor of $\sim2$ than
that obtained through fitting to the AMS-02 and Fermi-LAT data.

The Fermi-LAT observations of dwarf spheroidal galaxies
\cite{2011PhRvL.107x1302A,2011PhRvL.107x1303G,2013JCAP...03..018S,
2015PhRvL.115w1301A,2016PhRvD..93d3518L} and the Planck observations of
CMB anisotropies \cite{2016A&A...594A..13P} give effective and robust
constraints on the annihilating DM models. For the decaying DM, the
extragalactic $\gamma$-ray background (EGRB) is especially suitable
to constrain the DM model parameters \cite{2010JCAP...01..023C,
2010JCAP...07..008H,2012PhRvD..86h3506C,2017JCAP...03..041C}. We use
the {\tt LikeDM} code \cite{2017CoPhC.213..252H} to calculate the
constraints from Fermi-LAT observations of a population of dwarf spheroidal
galaxies (for annihilating DM models) and the EGRB \cite{2015ApJ...799...86A}
(for decaying DM models). In Fig. \ref{fig:sv-tau} we overplot the
constraints on the annihilating (left) and decaying (right) DM model
parameters on the $m_{\chi}-\sv$ or $m_{\chi}-\tau$ plane.
For annihilating DM models, we find that the channels with tauons will
produce too many photons and exceed the upper limits set by $\gamma$-ray
observations of dwarf galaxies. The channels with $e$ or $\mu$ final
state particles can survive the constraints. However, when compared with
the upper limits set by CMB (shaded region, with an energy deposition
efficiency of $f_{\rm eff}\sim 0.15-0.7$; \cite{2009PhRvD..80d3526S}),
none of these channels can survive. For decaying DM models, the recent 
EGRB data also tend to severely constrain all the models to explain the 
CR lepton data. Compared with recent studies \cite{2017ChPhC..41d5104L,
2017JCAP...03..041C} which showed that decaying DM models with $e^+e^-$
or $\mu^+\mu^-$ channels could (partially) survive the EGRB constraints,
the different conclusion obtained here is primarily due to that the
DAMPE data push the DM particle mass to the heavier end where the
constraints are stronger. These results suggest that to account for the
electron/positron excesses in the framework of DM, more tuning beyond
the current simplified DM models is required.

\subsection{Physical origin of the background electron spectrum}

The fittings reveal that the energy spectrum of background electrons
has a first break (softening) at several GeV, a second break (hardening)
at $O(100)$ GeV, and a cutoff at $\sim$TeV. The GeV break may be due to
the ion-neutral collisions around shocks which modifies (steepens) the
accelerated particle spectrum \cite{2011NatCo...2E.194M}. The spectral
hardening and cutoff may have a common origin, i.e., the breakdown 
of continuous source distribution and imprint of nearby source(s)
\cite{1995A&A...294L..41A,2011APh....35..211L,2014JCAP...04..006D}.
As a rough estimate, the total supernova rate in a volume within 1
kpc from the Earth, which is the effective transport range of TeV
electrons, is about $(1/15)^2\times10^{-2}\sim5\times10^{-5}$ yr$^{-1}$
assuming a total rate of $10^{-2}$ yr$^{-1}$ in the Milky Way.
The number of supernovae within the cooling time of $\sim3\times10^5$
yr is about 10. Observationally we indeed find roughly such a number
of nearby and fresh pulsars and/or SNRs \cite{2010A&A...524A..51D}.
Therefore the discreteness of those sources should be important in
regulating the TeV electron spectrum.

Alternatively, the break might be due to the cutoff of acceleration
electron spectrum at the source. This scenario is similar to the
``poly-gonato'' model to explain the knee of the total CR spectra
\cite{2003APh....19..193H}, i.e., the sum of sources with different
cutoff energies gives rise to the spectral break. Observationally,
however, we do find that many sources can actually accelerate electrons
to energies well beyond TeV \cite{2013FrPhy...8..714R}. Therefore, a
fine tuning of the source luminosity function of different cutoff
energies may be needed to explain the data.

\section{Interpretations of the peak feature at $\sim 1.4$ TeV}

The tentative spectral peak needs a very unusual intrinsic spectrum. It also
imposes a very stringent constraint on the distance of the source, because
the cooling would effectively smooth out the spectral features. For continuous
injection, the cooled electron spectrum should be $\propto E^{-2}$ at high
energies ($>$ a few tens of GeV where synchrotron and ICS coolings dominate),
which is too soft to be consistent with the data. The characteristic cooling
timescale of the electrons can be estimated as (with Eq. (\ref{be}), it is
straightforward to show that the first two terms are sub-dominant for the
cooling of TeV electrons)
\begin{equation}
\tau_{\rm cool} \sim 3\times10^{5}~{\rm yr}~\left(\frac{u_{\rm tot}}
{\rm eV~cm^{-3}}\right)^{-1}\left(\frac{E}{\rm TeV}\right)^{-1},
\end{equation}
where $u_{\rm tot}$ is the total energy density of the interstellar
radiation field and the magnetic field. The travel distance of these
electrons is limited to
\begin{equation}
\lambda\sim 0.8~{\rm kpc}~\left(\frac{D_0}{10^{28}~{\rm cm^{2}~s^{-1}}}
\right)^{1/2} \left(\frac{u_{\rm tot}}{\rm eV~cm^{-3}}\right)^{-1/2}
\left(\frac{E}{\rm TeV}\right)^{-1/3}.
\end{equation}
Therefore the new electron source should be relatively nearby.

\subsection{Astrophysical interpretations of the DAMPE peak}

\subsubsection{Energetics}

We consider instantaneous injection of electrons/positrons from the source.
In this case a source locating at a distance of $R_{\rm s}$ should
have a lifetime $\tau>\tau_{\rm s}$, where
\begin{equation}
\tau_{\rm s} \sim R_{\rm s}^{2}/4D(E) \sim 8~{\rm kyr}~
\left(\frac{R_{\rm s}}{0.1~{\rm kpc}}\right)^{2}\left(\frac{D(E)}
{10^{29}~{\rm cm^{2}~s^{-1}}}\right)^{-1}.
\end{equation}
The total energy of the source released can be roughly estimate as
\begin{eqnarray}
\varepsilon_{\rm tot} & \sim & {4\pi\over 3}\left(2\sqrt{D(E)\tau_{\rm s}}
\right)^{3}w_{\rm e} \nonumber\\
& \sim & 10^{46}~{\rm erg}~\left(\frac{D(E)}
{10^{29}~{\rm cm^{2}~s^{-1}}}\right)^{3/2}\left(\frac{\tau_{\rm s}}
{10^5~{\rm yr}}\right)^{3/2}\nonumber\\
& \times & \left({w_{\rm e}\over 1.2\times10^{-18}~{\rm erg~cm^{-3}}}\right),
\end{eqnarray}
where $w_e$ is the energy density of the TeV peak. Clearly, for
$\tau_{\rm s}\sim \tau_{\rm cool}\sim 3\times10^{5}~{\rm yr}$ and
$D({\rm TeV}) \sim 3.3\times10^{29}$ cm$^2$ s$^{-1}$, we have
$\varepsilon_{\rm tot} \sim 3\times10^{47}~{\rm erg}$, which is possible
for a pulsar with a rotation period shorter than $\sim 0.1~{\rm s}$.

To account for the DAMPE peak, the age of the source needs to be
relatively young (e.g., $\lesssim10^5$ yr) and the distance needs to be
close enough (e.g., $\lesssim1$ kpc). A few sources may satisfy these
conditions, such as Geminga, Monogem, Vela, Loop I, Cygnus Loop and so on
\cite{2010IJMPD..19.2011F,2012CEJPh..10....1P}.

\subsubsection{Injection energy spectrum}

While typical cutoff power-law spectrum from such astrophysical source(s)
could be able to explain the observed electron/positron excesses, the peak
structure in DAMPE spectrum requires quasi-monochromatic injection of
particles from the sources. The cold, ultra-relativistic $e^+e^-$ plasma
wind from pulsars located in the local bubble provides a possible
source of such electrons and positrons \cite{1984ApJ...283..694K,
2012Natur.482..507A,2012A&A...547A.114A}. The Lorentz factor of the
bulk flow of pulsar winds is suggested to be $\sim10^6$, which just
corresponds to the energy of the DAMPE peak. Furthermore, the density
cavity of the local bubble, in which our solar system lies, makes pulsar
winds less likely to form PWN, and the winds can easily escape and
transport to the Earth without modification of the energy spectrum.
Alternatively, Ref. \cite{2016arXiv161108384J} proposed a scenario of
the interaction between electrons and a kind of hypothetical particles
to produce quasi-monochromatic spectrum of electrons.

\begin{figure*}[!htb]
\includegraphics[width=0.45\textwidth]{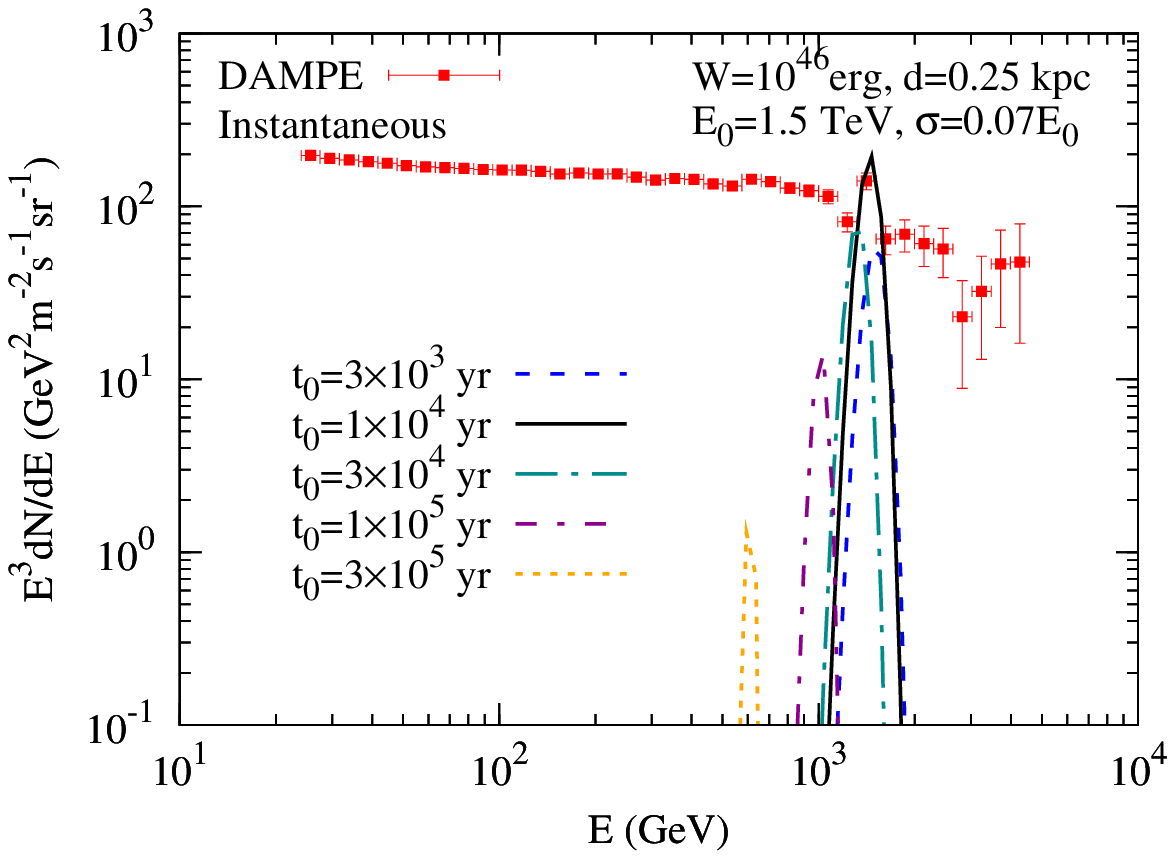}
\includegraphics[width=0.45\textwidth]{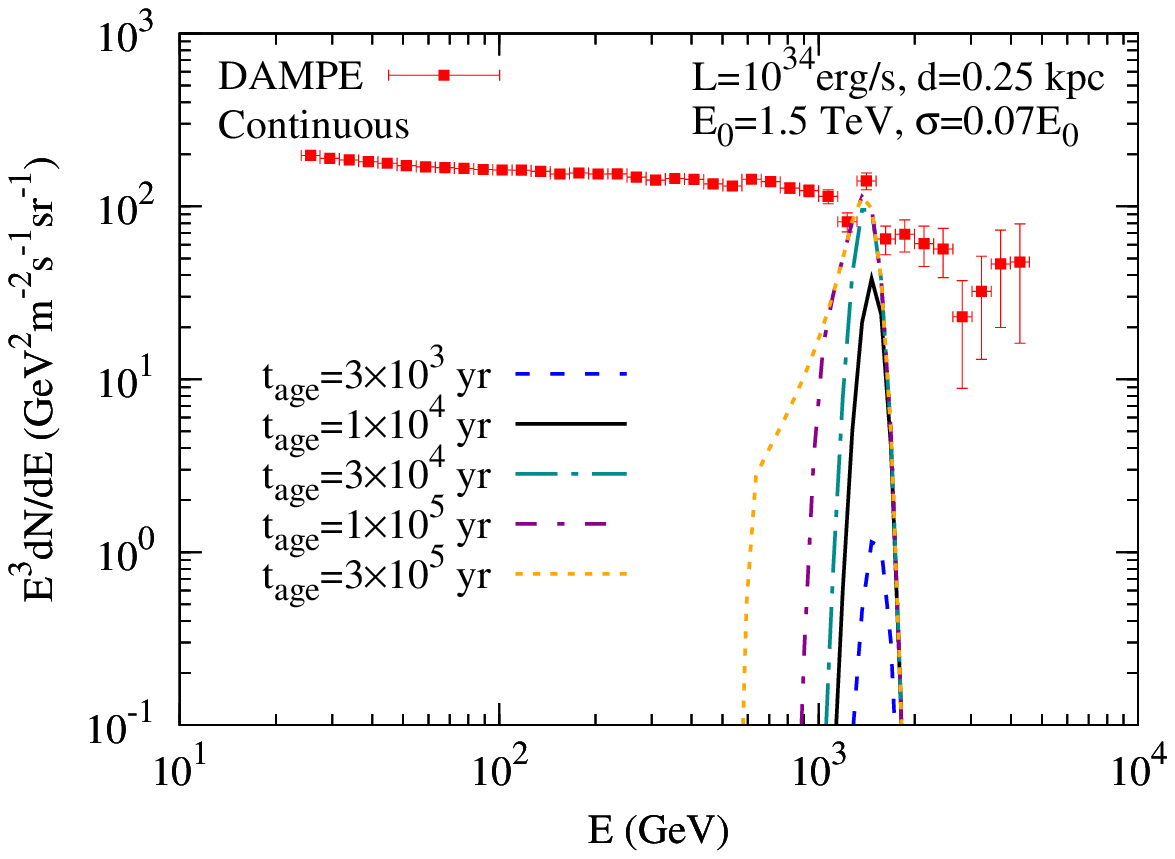}
\caption{Fluxes of total $e^++e^-$ from relativistic winds of a nearby
pulsar. The left panel is for instantaneous injection, and the right one
is for continuous injection. The injection spectrum is assumed to be
Gaussian with mean energy of $E_0=1.5$ TeV and width of $\sigma=0.07~E_0$.
\label{fig:psr_inst_cont}}
\end{figure*}

We assume a Gaussian spectrum to approximate the quasi-monochromatic
injection spectrum of electrons/positrons. Fig. \ref{fig:psr_inst_cont}
shows the expected $e^++e^-$ fluxes for instantaneous (left) and
continuous (right) injection of such relativistic $e^+e^-$ winds from
a nearby pulsar with a distance of $d=0.25$ kpc. The central energy
is assumed to be $E_0=1.5$ TeV, and the width is $\sigma=0.07~E_0$.
Different lines represent different injection time.

For instantaneous injection, two effects are clearly shown. First, we can
see a remarkable cooling effect on the spectrum. The earlier the injection,
the lower the cutoff energy. Second, an earlier injection also means a
larger diffusion length, and hence a lower flux. We can simply
estimate the diffusion length as $\lambda\sim2\sqrt{Dt}\approx 0.2~
{\rm kpc}~(E/{\rm TeV})^{1/6}(t/10^4\,{\rm yr})^{1/2}$. As long as
$t\gtrsim 10^4$ yr, we have $\lambda\gtrsim 0.2$ kpc for TeV electrons,
and hence $F\propto \lambda^{-3}\propto t^{-3/2}$. To match the data,
the required injection energy of the pulsar is about $10^{46}-10^{48}$
erg, depending on the injection time. Such a value is, however, a little
bit lower than that as expected from a typical pulsar
\cite{2009PhRvD..80f3005M,2013PhRvD..88b3001Y}.

For continuous injection, we find that with the increase of the integral
time, the high energy flux keeps on increasing until a certain level when
an equillbrium between injection and cooling is reached. Furthermore,
there are more low energy particles for longer injection time. This is
because for lower energy particles more time is needed for the injected
high energy ones to cool down. In the continuous injection scenario,
the spectrum is typically broader than that of instantaneous injection,
unless the source is fresh enough that cooling is unimportant. The
luminosity of the source, $\sim10^{34}-10^{36}$ erg s$^{-1}$, is again
relatively low compared with that of a typical pulsar.

\subsubsection{``Realistic'' pulsar model to account for the
electron/positron data}

\begin{figure*}[!htb]
\includegraphics[width=0.45\textwidth]{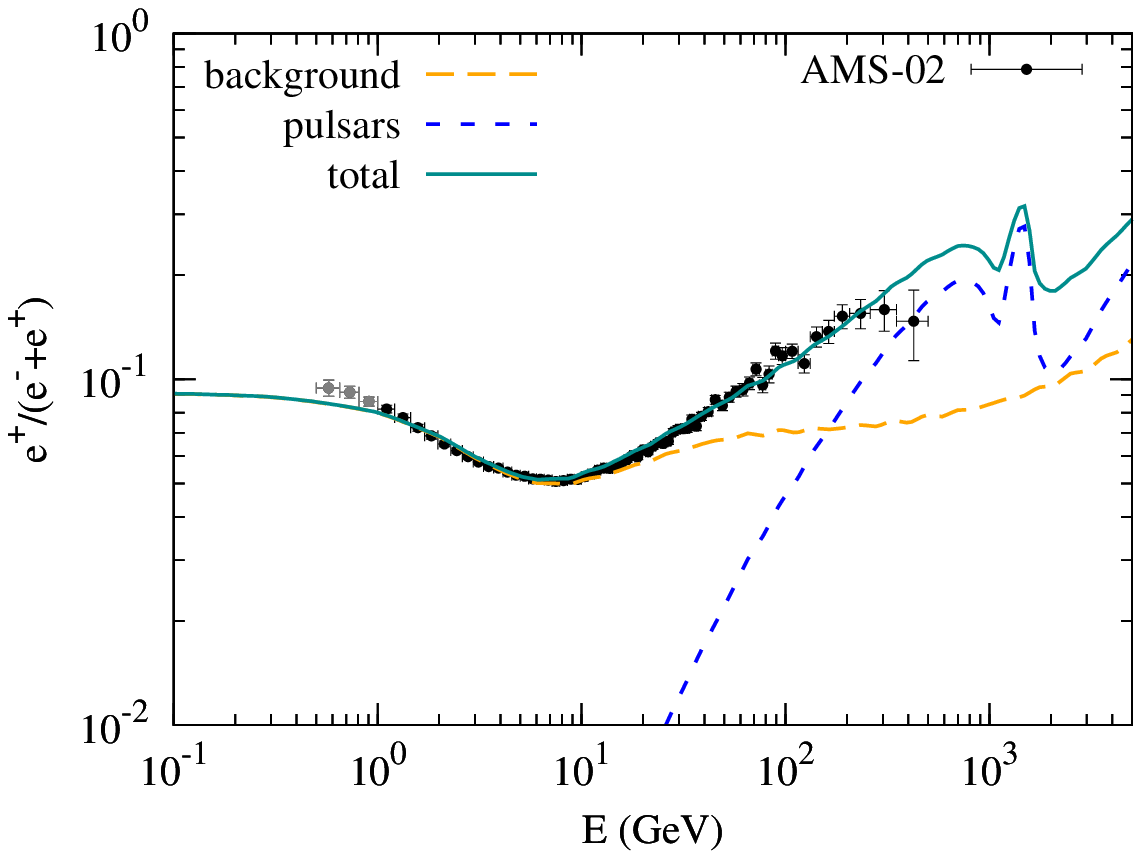}
\includegraphics[width=0.45\textwidth]{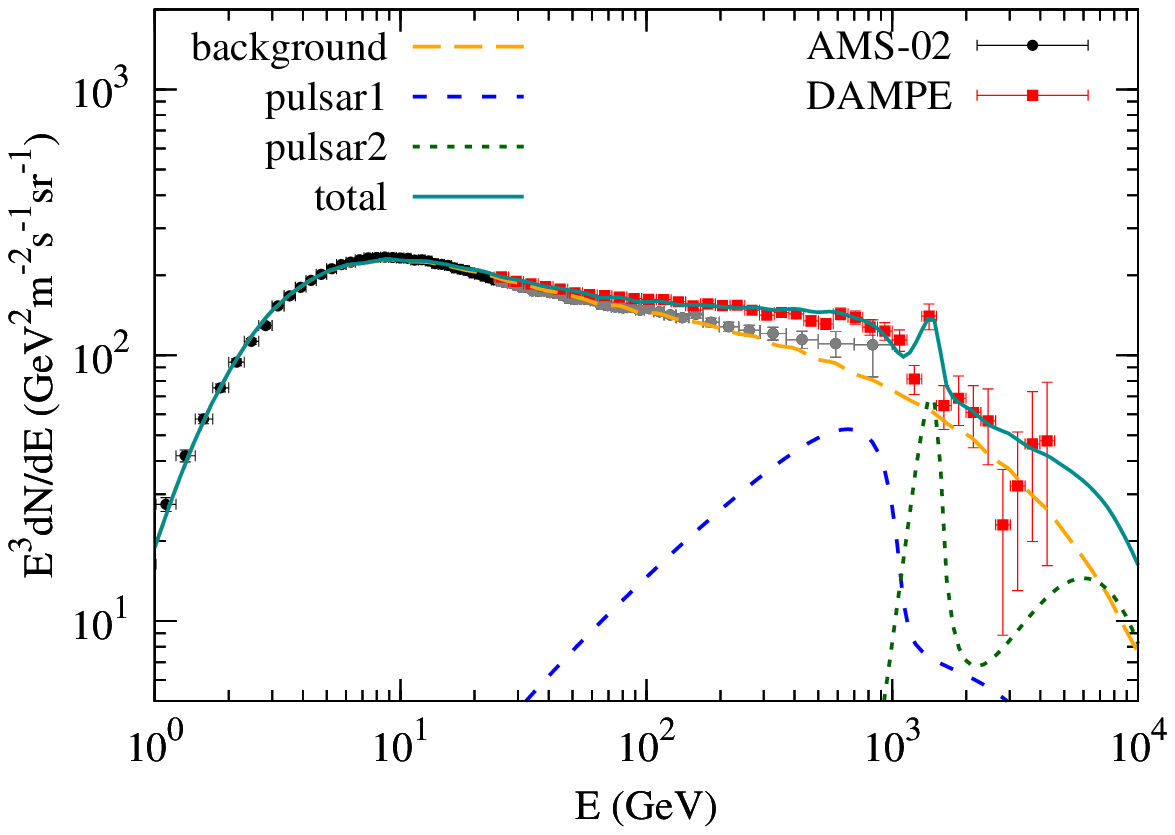}
\caption{The positron fraction (left) and total $e^++e^-$ fluxes (right)
for the model with two nearby pulsars. The background is the same as
that in Fig. \ref{fig:psr}.
\label{fig:psr_real}}
\end{figure*}

For a realistic pulsar model, the spin-down luminosity decays as $t^{-2}$
after the characteristic decay time $\tau_{\rm dec}$ which is typically
$10^3-10^4$ yr \cite{2009PhRvD..80f3005M,2012CEJPh..10....1P}.
Furthermore, the quasi-monochromatic injection spectrum may be too ideal.
In reality, a relativistic Maxwellian distribution, $f(E)\propto
E^2\exp(-E/\Theta)$ with $\Theta$ being the electron temperature, may be
more proper to describe the energy spectrum of the cold, untra-relativistic
$e^+e^-$ wind. We show in Fig. \ref{fig:psr_real} an illustration of the
positron fraction (left) and total $e^++e^-$ spectrum (right) from
two such ``realistic'' pulsars. The injection luminosity is assumed to be
a time-dependent form, $L=L_0\left(1+t/\tau_{\rm dec}\right)^{-2}$.
For pulsar 1, the injection electron/positron spectrum is assumed to be a
cutoff power-law form, $f(E)\propto E^{-\alpha}\exp(-E/E_{\rm max})$, which
could be due to acceleration and/or cooling in the nebula. For pulsar 2,
we assume a Maxwellian distribution of the injection wind particles.
The model parameters are tabulated in Table \ref{table:psr}. We can see
in Fig. \ref{fig:psr_real} that there are high energy tails of the pulsar
components, which are due to late time injection (see e.g.,
\cite{1995A&A...294L..41A}). The main peaks come from the early time
injection ($t<\tau_{\rm dec}$).

\begin{table}[!htb]
\caption {Parameters of the two pulsars as shown in Fig. \ref{fig:psr_real}.}
\begin{tabular}{cccccccc}
\hline \hline
 & $d$ & $L_0$ & $t_{\rm age}$ & $\tau_{\rm dec}$ & $\alpha$ & $E_{\rm max}$ ($\Theta$) \\
 & (kpc) & (erg s$^{-1}$) & (kyr) & (kyr) &  & (TeV) \\
\hline
pulsar 1 & 0.25 & $5.3\times10^{37}$ & 260 & 3.0 & 1.7 & 3.0 \\
pulsar 2 & 0.25 & $0.9\times10^{37}$ & 180 & 3.0 & $-2.0$ & 2.0 \\
\hline \hline
\end{tabular}
\label{table:psr}
\end{table}

Pulsar 1 is employed to account for the positron and electron excesses
below TeV. It may be Geminga- or Monogem-like (e.g., \cite{2009JCAP...01..025H,
2009PhRvL.103e1101Y,2013ApJ...772...18L,2013PhRvD..88b3001Y}). The initial
spin-down luminosity is estimated to be about $5\times10^{37}$ erg s$^{-1}$,
which is smaller by a factor of $\sim10$ than that of the crab pulsar.
This value is not unusual, since the crab pulsar is among the fastest
rotating family of normal pulsars \cite{2006ApJ...643..332F}. It is also
possible that the ensemble of a large population of pulsars in the Milky
Way plays the role of pulsar 1 \cite{2009PhRvD..80f3005M,2013PhRvD..88b3001Y}.
In such a case, wiggles may be seen in the spectrum \cite{2009PhRvD..80f3005M}
(see, however, \cite{2010ApJ...710..958K}).

Pulsar 2 is introduced to explain the DAMPE peak. Its properties are more
unique. First, the $e^+e^-$ winds should be directly injected into the
interstellar space without effective acceleration by the SNR/PWN. This
implies that either the pulsar moves fast away from the SNR, or the
supernova explodes in a density cavity (e.g., the local bubble) and the
ejecta expands fast without significant deceleration. Second, moderate
cooling of electrons is necessary to form a peak of the spectrum,
otherwise the spectrum would be too broad for Maxwellian injection.
This requires that the pulsar should be middle-aged (mature). On the
other hand, the pulsar's age should not be too large to make most of
the electrons cool down to energies lower than TeV. Therefore the age
of pulsar 2 is constrained to be about $10^5$ yr. The initial spin-down
luminosity of pulsar 2 is about $10^{37}$ erg s$^{-1}$, two orders of
magnitude lower than the current crab pulsar's spin-down luminosity.
This luminosity is also reasonable, suggesting a rotation period of
$\sim 0.15~{\rm s}~(L/10^{37}~{\rm erg~s^{-1}})^{1/4}~(B/10^{13}~
{\rm G})^{-1/2}$ for a neutron star radius $\sim 10$ km. In Ref.
\cite{2004ApJ...617L.139V} it was shown that $\sim40\%$ normal pulsars are
born with periods of $0.1-0.5$ s, about an order of magnitude longer than
that of the crab pulsar.

For the pulsars assumed in the above discussion, the surface magnetic field
can be estimated as $B=8.6\times10^{12}(P/0.1~{\rm s})~(\tau_{\rm dec}/
10^4~{\rm yr})^{-1/2}$ G $\sim10^{13}$ G \cite{2010ApJ...710..958K}.
Therefore mildly magnetized pulsar is needed to explain the data.

We conclude that a {\it middle-aged, relatively slowly-rotated, mildly
magnetized, isolate pulsar in the local bubble} is a possible candidate
source of the DAMPE peak. There may be more than one pulsars like pulsar 
2 in the local environment. In that case we may expect to see more such 
peaks in the electron spectrum. The future measurement of the electron 
spectrum to higher energies by DAMPE with high energy resolution will 
critically test this scenario.

\subsection{DM annihilation interpretations of the DAMPE peak}

\subsubsection{General consideration}

Suppose the source injects monoenergetic electrons at a continuous rate
$\dot{Q}$, the radial energy density distribution of electrons can be
approximately as \cite{2016Natur.531..476H}
\begin{equation}
w_{\rm e}(R)={\dot{Q}\over 4\pi D(E)R}{\rm erfc}(R/\lambda).
\end{equation}
Note that for monoenergetic electrons the above equation holds only for
$R\ll \lambda$; otherwise the cooling effect can not be ignored any more.
Under such a condition we have $w_{\rm e}(R)\approx \dot{Q}/[4\pi D(E)R]$,
and the source injection power is
\begin{eqnarray}
\dot{Q}&\approx& 5\times 10^{32}~{\rm erg~s^{-1}}~\left({R\over
0.1~{\rm kpc}}\right)\left({D(E)\over 10^{29}~{\rm cm^{2}~s^{-1}}}\right)
\nonumber\\
&\times &\left({w_{\rm e}\over 1.2\times10^{-18}~{\rm erg~cm^{-3}}}\right).
\end{eqnarray}

In the case of a DM sub-halo with a dense core whose size is $\delta$,
the energy density of the DM particles should be (assuming $m_{\chi}
\approx1.5$ TeV)
\begin{eqnarray}
\rho_{\chi} & \sim & 0.4~{\rm TeV~cm^{-3}}~\left({\dot{Q}\over 5\times
10^{32}~{\rm erg~s^{-1}}}\right)^{1/2}\nonumber\\
& \times & \left[{\langle \sigma v \rangle_{\chi\chi\rightarrow e^{+}e^{-}}
\over 3\times 10^{-26}~{\rm cm^3~s^{-1}}}\right]^{-1/2}\left({\delta \over
10~{\rm pc}}\right)^{-3/2},
\end{eqnarray}
where $\langle \sigma v \rangle_{\chi\chi\rightarrow e^{+}e^{-}}$ is the
velocity-averaged annihilation cross section. This density is about
1000 times higher than that of the widely-taken local DM energy density.
Interestingly there is another possibility that the Sun is actually in a
DM-density-enhanced region. Suppose such a region has a size of $\sim 100$
pc, we have $\rho_{\chi}\sim 0.012~{\rm TeV~cm^{-3}}$, about 30 times
higher than the canonical local DM energy density. Such a density
enhancement seems to be not in tension with other observations
\cite{2014PhLB..728...58H}. The total mass of the DM sub-halo or
density-enhanced region reads
\begin{eqnarray}
M_{\chi} & \sim & 4\times 10^{4}~M_\odot~\left({\dot{Q}\over 5\times
10^{32}~{\rm erg~s^{-1}}}\right)^{1/2}\nonumber\\
& \times & \left[{\langle \sigma v \rangle_{\chi\chi
\rightarrow e^{+}e^{-}}\over 3\times 10^{-26}~{\rm cm^3~s^{-1}}}\right]
^{-1/2}~\left({\delta \over 10~{\rm pc}}\right)^{3/2}.
\end{eqnarray}

The energy spectrum of electrons at production depends on the mass of the
DM particle and the annihilation final state. For the quark final state,
the electrons are produced through the hadronization of quarks, which
results in a very soft spectrum. The spectrum is significantly harder
for the leptonic and gauge bosonic final states due to the non-zero
decaying branching ratios to direct production of electrons. The hardest
spectrum comes from the direct annihilation into a pair of $e^+e^-$,
which gives nearly monochromatic electrons (and positrons). The peak
structure of the DAMPE data requires a branching ratio to $e^+e^-$
final state, and a local origin of the DM annihilation. These requirements
suggest the scenario of a nearby DM clump \cite{2009PhRvD..79j3513H,
2009PhRvD..80c5023B} or a local DM density enhancement
\cite{2014PhLB..728...58H}.


In order to compare with the data, we also need to include the
``background'' electrons/positrons. The continuous $e^++e^-$ fluxes that
derived through fitting to the wide-band data in Sec. III are assumed to
be the ``background'' here (similar approach has been used to search
for potential prominent spectral features; \cite{2013PhRvL.111q1101B,
2013PhRvL.110d1302H,2014PhRvD..89f3539I}). Specifically, the fluxes 
shown in Fig. \ref{fig:psr} are adopted.

\subsubsection{Nearby DM clump}

The structures of matter in the universe are hierarchical. The N-body
simulations of the cold DM (CDM) structure formation reveals a large
amount of subhaloes in the Galactic DM halo \cite{2004MNRAS.355..819G,
2008MNRAS.391.1685S}, with the smallest subhalo mass as light as the
Earth \cite{2005Natur.433..389D}. The DM subhaloes were exployed to
account for the ``boost factor'' required to explain the HEAT positron
excess \cite{2007JCAP...05..001Y}. More detailed calculation based on
the N-body simulation results showed that the ``boost factor'' from DM
clumpiness was usually negligible \cite{2008A&A...479..427L}.
Alternatively, the local DM subhalo(es) were able to play the role of
an effective ``boost factor'' if they are close enough to the Earth
\cite{2007MNRAS.374..455C,2009PhRvD..79j3513H}. In addition, this
scenario can give harder positron spectrum than that of the average
of the whole Galaxy, which can fit the data with more conventional
annihilation channels of DM (e.g., gauge bosons and quarks). However,
the comparison of the required conditions with the numerical simulations
shows that the probability to have such a DM clump close and massive
enough is very low \cite{2007A&A...462..827L,2009PhRvD..80c5023B}.
Here we revisit this nearby DM clump model in light of the DAMPE data,
and investigate what condition is needed to account for the data.



\begin{figure*}[!htb]
\includegraphics[width=0.45\textwidth]{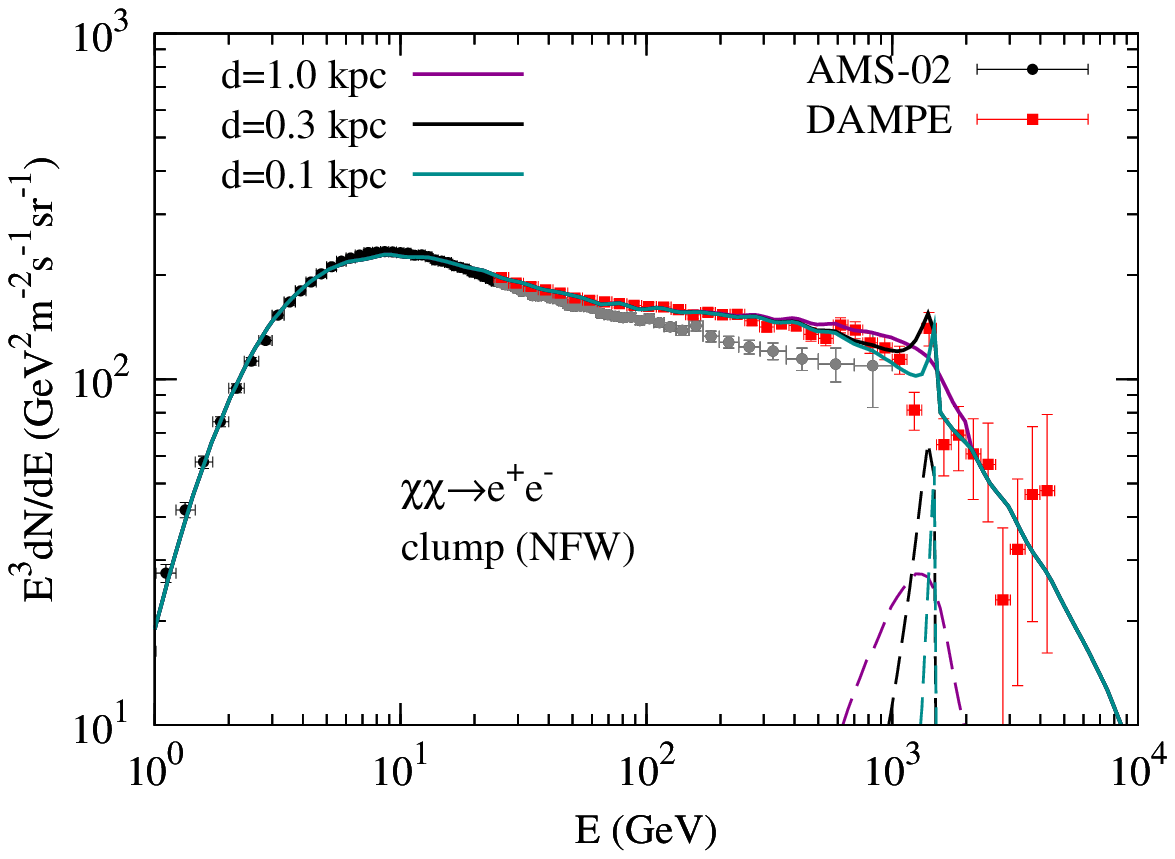}
\includegraphics[width=0.45\textwidth]{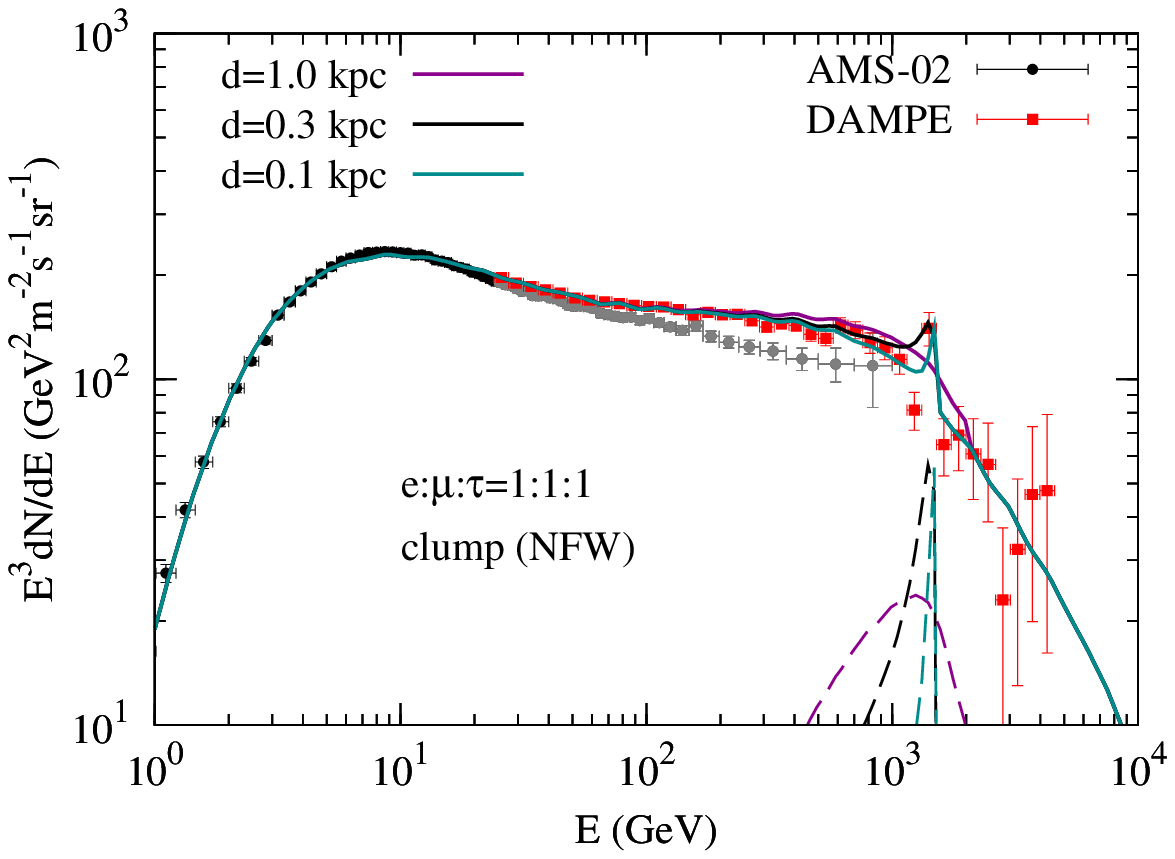}
\caption{Fluxes of the total $e^++e^-$, from the sum of the continuous
background and the DM annihilation from a nearby clump. The left panel
is for DM annihilation into a pair of $e^+e^-$, and the right one is
for DM annihilation into all flavor leptons with universal couplings.
Three distances of the clump, as labelled in the plot, are considered.
See Table \ref{table:clump_nfw} for the mass of the DM particle, and
the mass and annihilation luminosity of the clump.
\label{fig:elec_clump}}
\end{figure*}


\begin{table*}
\centering
\caption{Mass of the DM particle, and mass and luminosity of the DM
clump halo to fit the DAMPE data. The annihilation cross section is
assumed to be $3\times10^{-26}$ cm$^3$s$^{-1}$.}
\begin{tabular}{ccccccccccccccc}
\hline \hline
channel  & \multicolumn{3}{c}{1.0 kpc} & & \multicolumn{3}{c}{0.3 kpc} & & \multicolumn{3}{c}{0.1 kpc} \\
    \cline{2-4} \cline{6-8} \cline{10-12}
    &  $m_{\chi}$/TeV & $M_{\rm sub}$/M$_\odot$ & ${\mathcal L}$/GeV$^2$cm$^{-3}$ & & $m_{\chi}$/TeV & $M_{\rm sub}$/M$_\odot$ & ${\mathcal L}$/GeV$^2$cm$^{-3}$ & & $m_{\chi}$/TeV & $M_{\rm sub}$/M$_\odot$ & ${\mathcal L}$/GeV$^2$cm$^{-3}$ \\
\hline
$e^+e^-$ & 2.2 & $3.8\times10^{9}$ & $1.0\times10^{67}$ & & 1.5 & $8.0\times10^{7}$ & $3.8\times10^{65}$ & & 1.5 & $5.0\times10^{6}$ & $3.5\times10^{64}$ \\
$e\mu\tau$ & 2.2 & $1.0\times10^{10}$ & $2.3\times10^{67}$ & & 1.5 & $2.6\times10^{8}$ & $1.0\times10^{66}$ & & 1.5 & $1.9\times10^{7}$ & $1.1\times10^{65}$ \\
\hline
\hline
\end{tabular}
\label{table:clump_nfw}
\end{table*}

We assume that the density profile of the DM clump is NFW. For a clump
close to the solar location, the strong tidal force of the Milky Way
will remove the outer material of the clump, leaving a more compact core.
We present the detailed determination of the mass distribution of a DM
clump when considering the tidal stripping in the Appendix. We further
assume that the annihilation cross section is $3\times10^{-26}$
cm$^3$s$^{-1}$, which corresponds to the value inferred from the thermal
production of the DM. Fig. \ref{fig:elec_clump} shows the resulting
$e^++e^-$ fluxes for a DM clump locating at 1.0, 0.3, and 0.1 kpc away
from the Earth. The left panel is for the $e^+e^-$ final state, and the
right one is for annihilation to all flavors of charged leptons with equal
branching ratios. The required DM particle mass, the total mass of the
clump, and its annihilation luminosity, ${\mathcal L}=\int\rho^2dV$,
are given in Table \ref{table:clump_nfw}.

We find that a DM clump as massive as $10^7-10^8$ M$_{\odot}$ with a
distance of $\sim0.1-0.3$ kpc can account for the DAMPE data if the
annihilation branching ratio to $e^+e^-$ is large enough.
We have tested that for channels other than $e^+e^-$, such as
$\mu^+\mu^-$, $\tau^+\tau^-$, and $W^+W^-$, no prominent peak feature
can be generated. The distance of the DM clump should not be larger
than a few hundred parsec so that the $\delta$-function like spectrum
of electrons/positrons will not be smoothed out by the cooling. These
two requirements are just consistent with what we expected in Sec.
IV. A. Note also that for a distance of $1.0$ kpc, the required clump
mass is about $10^{10}$ M$_{\odot}$, which is close to the maximum
mass of subhalos in A Milky Way size halo \cite{2008MNRAS.391.1685S}.
Such a case is thus disfavored.

We need further to check that the contribution from such a DM clump
dominates the Milky Way halo and other substructures so that the peak
structure would not be smoothed out. Since the average enhancement due to
substructures for charged particles is found to be small
\cite{2008A&A...479..427L}, we consider only the Milky Way halo. It turns
out that, for the $e^+e^-$ channel, $\sv=3\times10^{-26}$ cm$^3$s$^{-1}$
and $m_{\chi}=1.5$ TeV, the contribution from the DM annihilation in the
whole Milky Way halo is about two orders of magnitude lower than that
shown in Fig. \ref{fig:elec_clump} at $\sim$TeV, which is thus negligible
(this can also be seen from Fig. \ref{fig:sv-tau}).

\subsubsection{Enhanced local DM density}

Besides the gravitationally bound substructures, density fluctuations occur
and disappear continuously in the Milky Way halo, as illustrated by
numerical simulations \cite{2010PhRvD..81d3532K}. An over-density region
with $\rho/\rho_0<200\,({\rm kpc}/\delta)^2$, where $\rho_0$ is the average
local density and $\delta$ is the region size, is gravitationally
unbound \cite{2014PhLB..728...58H}. The density profile within the
over-density regions may be shallower than the gravitationally bound
substructures. It was proposed that a local DM density enhancement by
a factor of $\sim30-50$ within $R\sim0.5$ kpc of the Earth and annihilation
channels into gauge bosons or quarks could explain the observed positron
excesses \cite{2014PhLB..728...58H}.


The results of the enhanced local density scenario are very similar to
that of Fig. \ref{fig:elec_clump}.
To give the DAMPE peak, we need a considerable branching ratio to
monochromatic $e^+e^-$ in the final states and a not too large size
($s\lesssim0.3$ kpc) of the over-density region. If the annihilation
cross section is assumed to be $\sv=3\times10^{26}$ cm$^3$s$^{-1}$,
the required density is about $17-35$ times of the canonical local
density of $\rho_0=0.4$ GeV cm$^{-3}$ for the $e^+e^-$ channel.

\subsubsection{DM (sub-)structures from numerical simulations}

In the above two subsections, we have shown that either a nearby DM
clump or an enhanced local density is able to account for the DAMPE
electron spectrum. Here we check from the numerical simulations of the
DM structure formation to see whether the required conditions can be
satisfied.

\begin{figure}[!htb]
\includegraphics[width=0.45\textwidth]{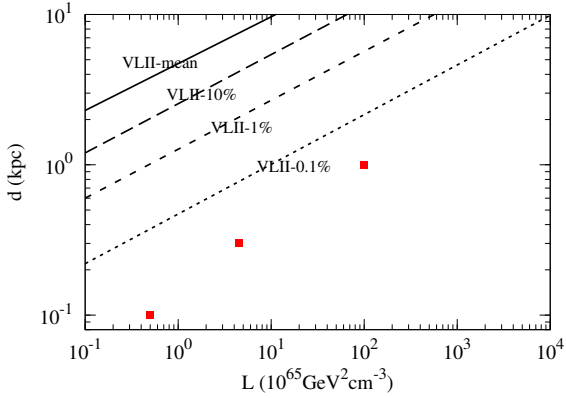}
\caption{Required annihilation luminosity ${\mathcal L}$ and distance
$d$ (red squares) from the Earth of the clump to fit the DAMPE data,
assuming $e^+e^-$ channel, compared with the probability distributions
(black lines) of finding a subhalo with ${\mathcal L}$ within $d$ inferred
from the VLII simulations \cite{2009PhRvD..80c5023B}.
\label{fig:prob_sub}}
\end{figure}

Fig. \ref{fig:prob_sub} shows the probability distributions of finding
a clump with annihilation luminosity ${\mathcal L}$ within distance $d$
from the Earth inferred from the Via Lactea II (VLII) simulations
\cite{2008Natur.454..735D}, as given in Ref. \cite{2009PhRvD..80c5023B}.
Red squares are the required values to account for the DAMPE electron
data, as discussed in Sec. IV. B. Here the annihilation channel is assumed
to be $e^+e^-$, and the cross section is fixed to be $3\times10^{-26}$
cm$^3$s$^{-1}$. It is shown that the probability of having a right clump
to explain the data is relatively low, $p<10^{-3}$. For the annihilation
to $e\mu\tau$ leptons with universal couplings, the required luminosity
${\mathcal L}$ is larger by a factor of 3, hence the situation is even
worse. If there is moderate enhancement of the annihilation cross section
from the particle physics, such as the non-thermal production of the DM
\cite{1999JHEP...12..003J,2001PhRvL..86..954L,2012PhRvD..86j3531Y},
the Sommerfeld enhancement mechanism (e.g., \cite{Sommerfeld1931,
2005PhRvD..71f3528H,2009PhRvD..79a5014A}) or Breit-Wigner type resonance of
the annihilation interaction \cite{1991PhRvD..43.3191G,1991NuPhB.360..145G},
the required luminosity can be lower and the probability becomes higher.
Alternatively, if there is mini-spike in the center of the DM clump,
the annihilation luminosity may be significantly enhanced and the required 
mass of the clump may be much smaller \cite{2005PhRvL..95a1301Z}.

As for the enhanced local density scenario, we consider the distribution
of density fluctuations. The density fluctuation was found to follow
roughly a log-normal distribution with a Gaussian width of $\ln\rho$ of
$\sim0.2$ \cite{2010PhRvD..81d3532K}. Therefore an over-density of $20$
times of the local average density indicates a $\sim15\sigma$ deviation
from the mean value. This probability, again, is very low. Note,
however, the distribution of density fluctuation has a power-law tail
at large $\rho$, due to substructures \cite{2010PhRvD..81d3532K}.
Therefore the actual probability to have an over-density region should
be higher than the above estimate. Furthermore, if there is enhancement
of the annihilation cross section the required over-density factor can
be smaller.

\subsubsection{Gamma-ray constraints}

The annihilation of DM will produce $\gamma$-ray photons accompanied with
electrons/positrons, from the internal bremsstrahlung process for the
charged fermion channel and/or the decay of the final state particles.
We check whether such $\gamma$-ray emission can be detectble or constrained
by the current observations.

\begin{figure*}[!htb]
\includegraphics[width=0.45\textwidth]{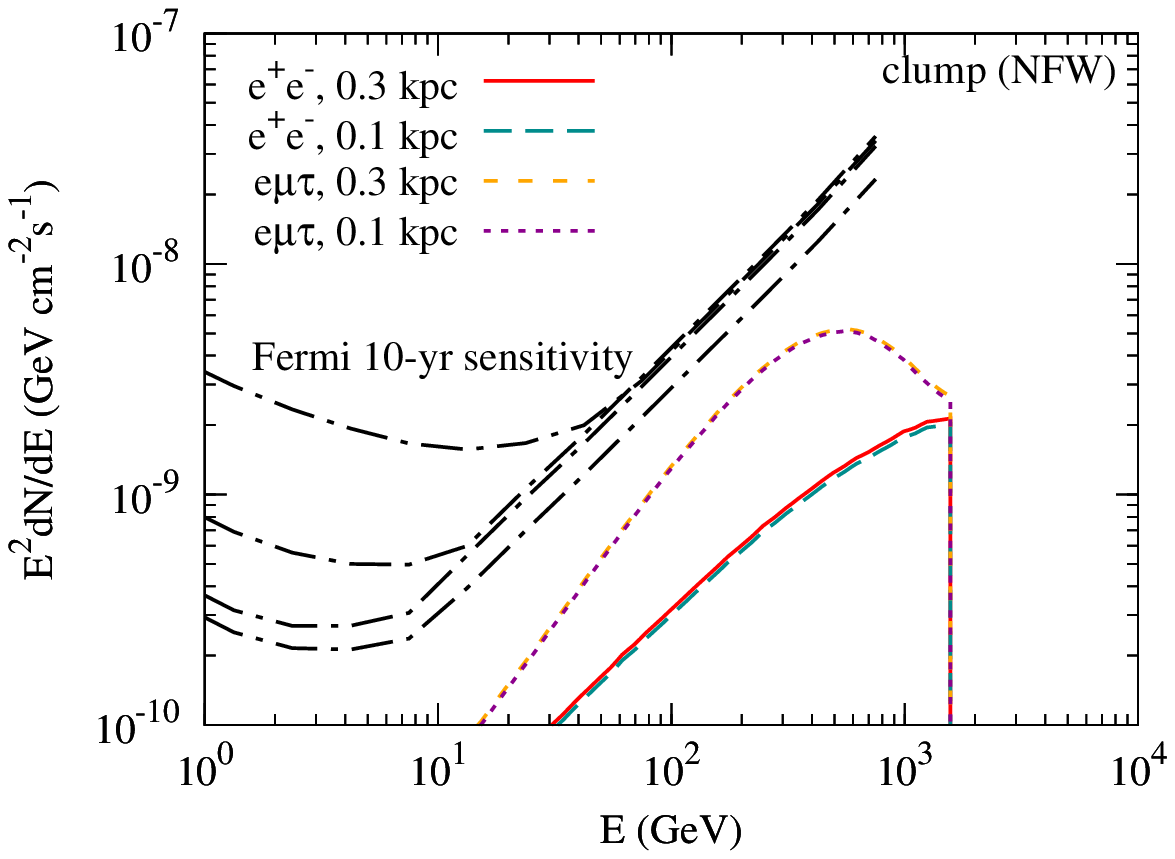}
\includegraphics[width=0.45\textwidth]{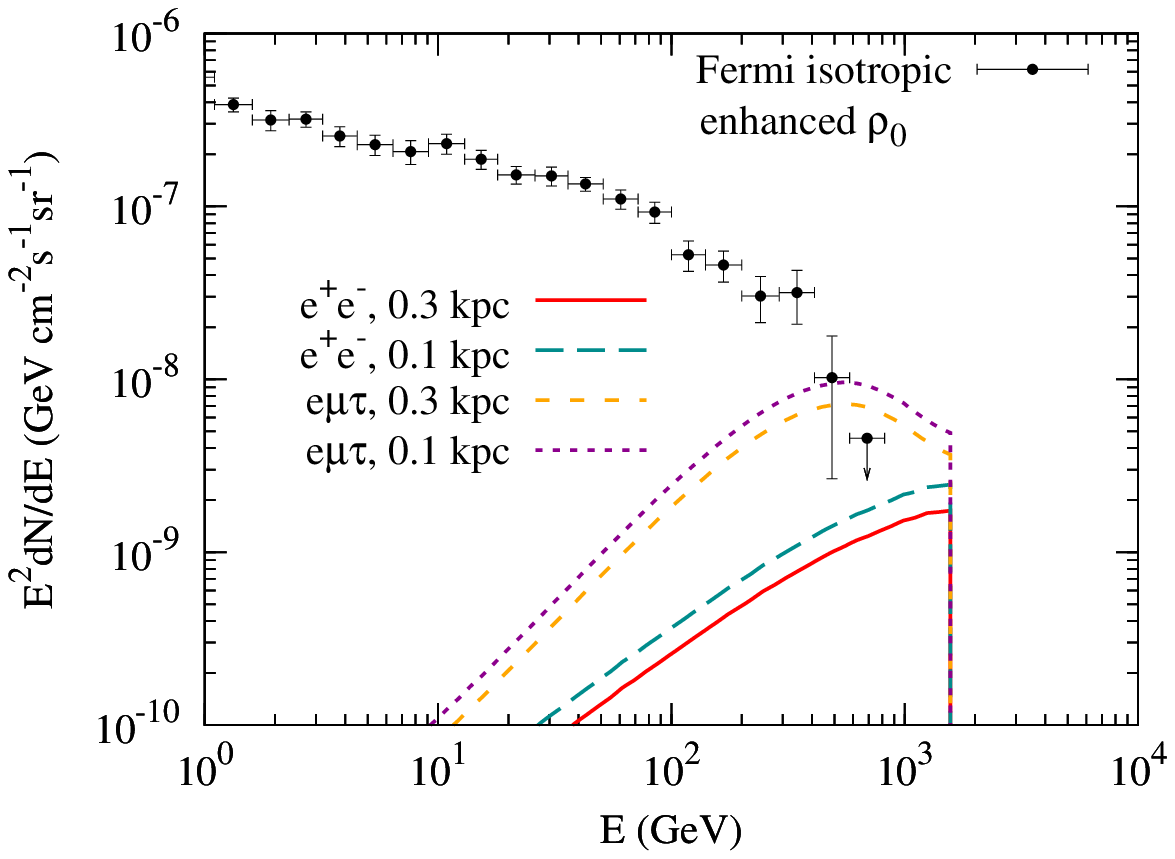}
\caption{Left: expected $\gamma$-ray fluxes from the annihilation of DM
in a nearby clump. The integral radius is chosen to be $1^{\circ}$. The
dot-dashed lines show the 10-year point source sensitivities of Fermi-LAT
with the Pass 8 instrument response for directions of $(l,b)=(0,0)$,
$(0,30^{\circ})$, $(0,90^{\circ})$, and $(120^{\circ},45^{\circ})$ from
top to bottom.
Right: expected $\gamma$-ray fluxes from the annihilation of DM in the
enhanced local density model, compared with the isotropic background
measurements by Fermi-LAT \cite{2015ApJ...799...86A}.
\label{fig:gamma_dm}}
\end{figure*}


We calculate the expected $\gamma$-ray fluxes from the DM annihilation
for the models which can potentially explain the DAMPE electron data,
i.e., with $e^+e^-$ or $e\mu\tau$ channels and a distance (for the clump)
or radius (for the enhanced local density scenario) of $\leq0.3$ kpc.
For the nearby clump scenario, the $\gamma$-ray emission is essentially
extended. For simplicity, we integrate the emission within $1^{\circ}$
radius, and compare them with the 10-year point source sensitivity of 
Fermi-LAT with the Pass 8 instrument 
response\footnote{http://www.slac.stanford.edu/exp/glast/groups/canda/lat\_Performance.htm}. 
For the enhanced local density scenario, the $\gamma$-ray emission is 
diffuse. Therefore we employ the Fermi-LAT isotropic background data as 
constraints \cite{2015ApJ...799...86A}. The results are shown in Fig. 
\ref{fig:gamma_dm}. We find that except for the annihilation to $e\mu\tau$ 
case of the enhanced DM density model which exceeds the highest point of 
the Fermi-LAT isotropic background marginally, other cases are not in 
conflict with Fermi-LAT observations.



\subsubsection{Particle model of DM}

In this sub-section we discuss possible DM particle models which are able to 
interpret the DAMPE peak, and the relevant constraints from the relic density.

{\it Lepton Portal DM model} ---
Following \cite{2014JHEP...08..153B}, we discuss the lepton portal DM models.
If DM particles are fermions, the interaction between DM and leptons can
be written as
\begin{equation} \label{LtdirS}
\mathcal{L}_{\rm fermion} \supset  \lambda_i \phi_i \bar{\chi}_L  e_R^i + h.c.,
\end{equation}
where $\chi$ denotes the DM particle, $e^i=e,\mu,\tau$ represents charged
leptons, and $\phi_i$ is the charged scalar mediator with unit lepton number.
In this model, we have only two parameters (the coupling strength $\lambda_i$
and the mass of the mediator $m_{\phi,i}$) for each flavor. To avoid the
decay of DM, $m_\chi$ should be smaller than the mass of the mediator.

If DM particles are Dirac fermions, the annihilation cross section is
\begin{eqnarray}
\label{eq12}
\frac{1}{2}\,(\sigma v)^{\chi\bar\chi}_{\rm{Dirac}} & = & \frac{1}{2}\left[
\frac{\lambda^4 m_\chi^2}{32\,\pi\, (m_\chi^2 + m_{\phi}^2)^2}\right. \\
& + & \left. v^2 \frac{\lambda^4\,m_\chi^2 \,( - \,5 m_\chi^4 \,-\,
18 m_\chi^2 m_{\phi}^2 + 11 m_{\phi}^4 ) }{768\,\pi\, (m_\chi^2 +
m_{\phi}^2)^4 } \right],\nonumber
\end{eqnarray}
where $v$ is the relative velocity between two DM particles and the factor
$1/2$ accounts for the difference between particle and anti-particle of
Dirac DM. Here we neglect the mass of leptons, and assume that
$\lambda_e = \lambda_{\mu}= \lambda_{\tau} \equiv \lambda$. Obviously,
the cross section is not velocity suppressed in this model.

\begin{figure*}[htb]
\includegraphics[width=0.45\textwidth]{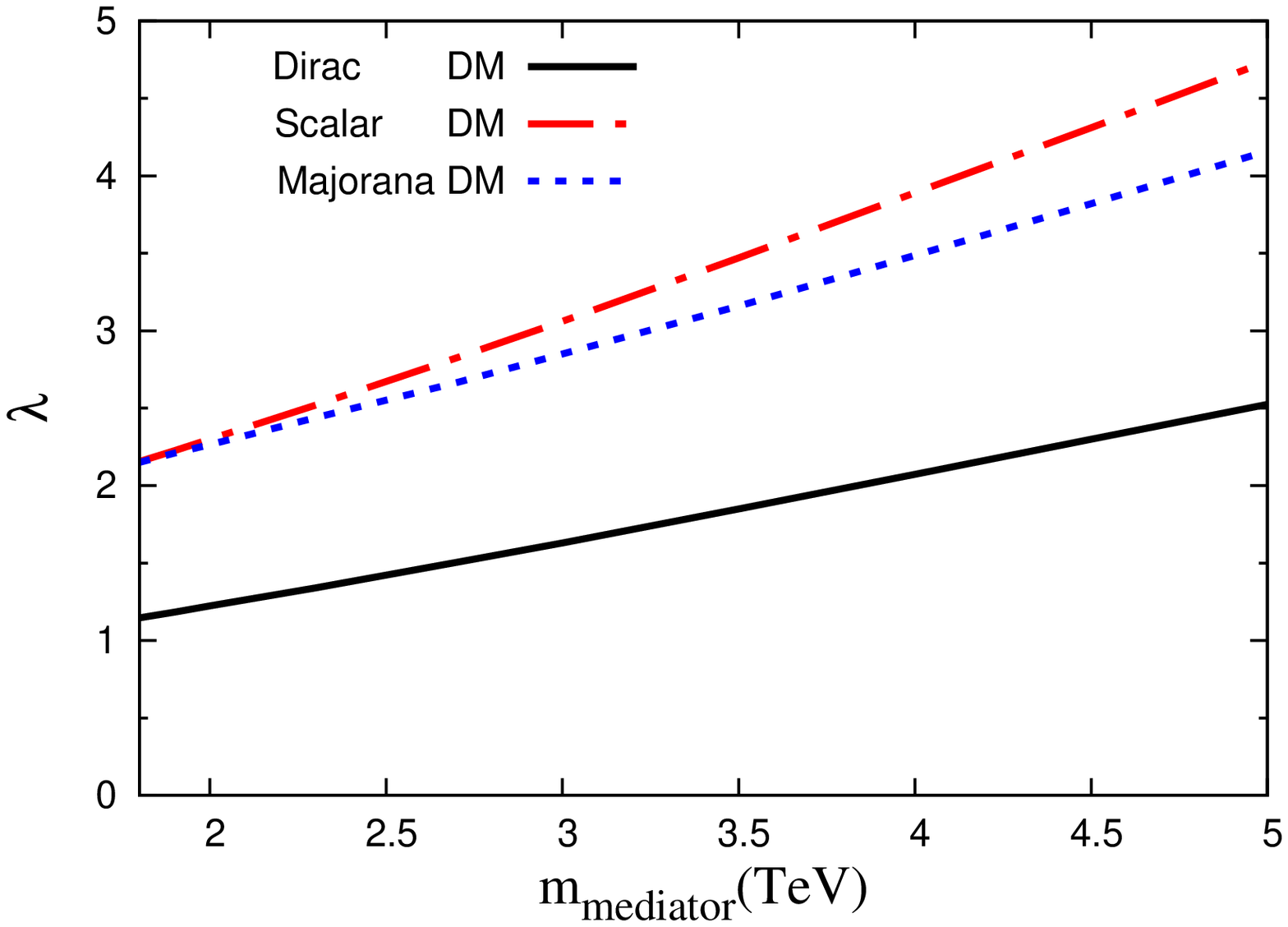}
\includegraphics[width=0.45\textwidth]{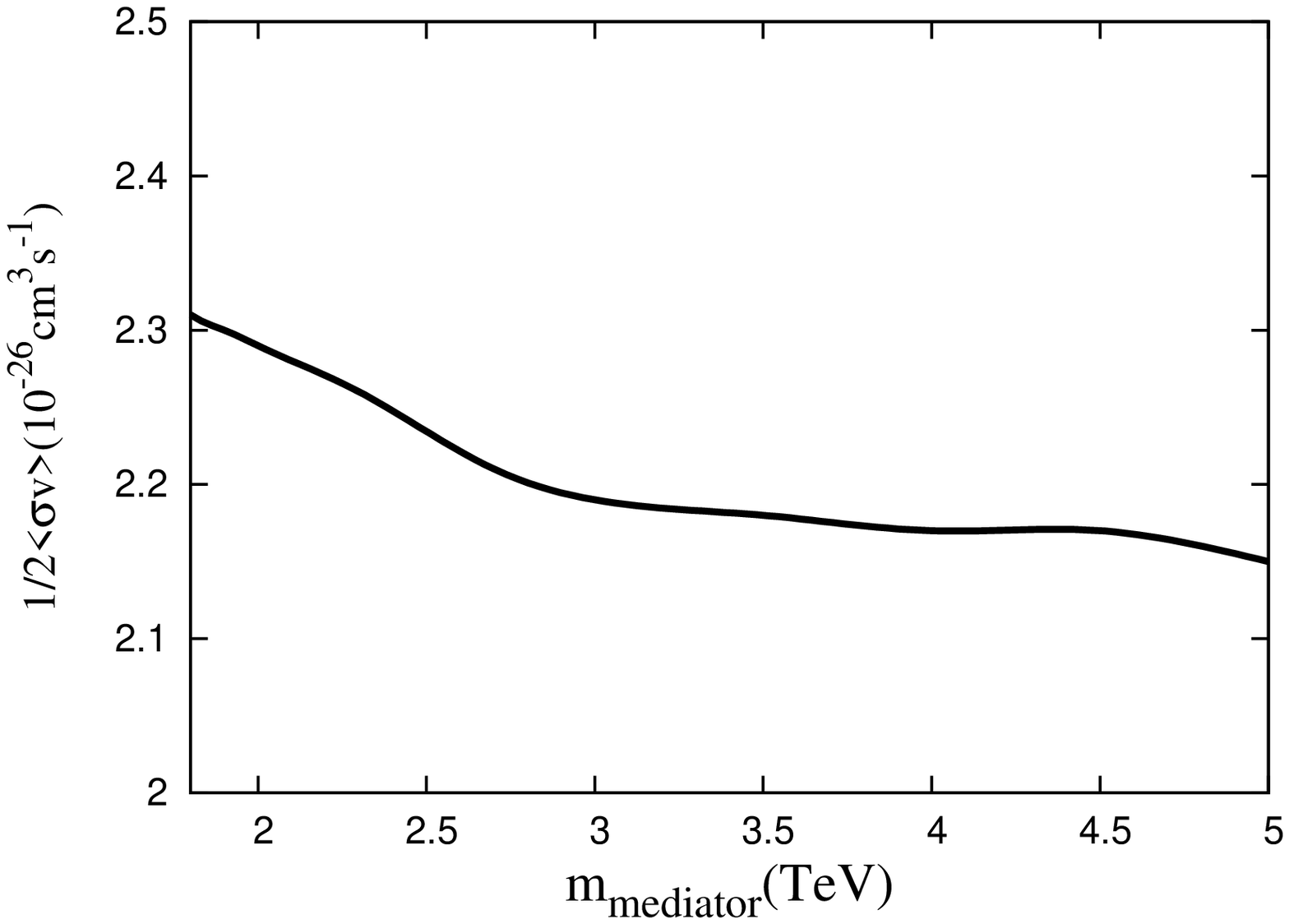}
\caption{The coupling constants $\lambda$ as functions of the mediator mass 
that give the correct relic density (left), and the corresponding 
annihilation cross section for the Dirac fermion model (right).}
\label{fig:md}
\end{figure*}

We calculated the relic density of DM using micrOMEGAs
\cite{2007CoPhC.176..367B}. The parameters that produce the correct DM
relic density, $\Omega_{DM}h^2=0.119$, for different values of $m_\phi$
are shown in the left panel of Fig.~\ref{fig:md}. The corresponding
annihilation cross section ($\langle \sigma v \rangle/2$) is shown in
the right panel of Fig.~\ref{fig:md} and the factor of 1/2 accounts for 
the fact that Dirac DM consists of both a particle and an anti-particle.
In the previous sub-section, we set $\sv=3\times10^{-26}$ cm$^3$s$^{-1}$ 
and get the corresponding annihilation luminosity $L=\int \rho^2 dV$. 
From Fig.~\ref{fig:md}, we can see that there is a slight deviation of the 
DM annihilation cross section from the canonical value $3\times10^{-26}$ 
cm$^3$s$^{-1}$. So the annihilation luminosity should be
\begin{equation}
L=L^\prime\times\frac{3\times 10^{-26}{\rm cm^3s^{-1}}}{1/2\langle 
\sigma v \rangle}. \nonumber
\end{equation}
Here $L^\prime$ is the annihilation luminosity presented in Table
\ref{table:clump_nfw}.

For the Majorana DM case, the annihilation cross section is
\begin{equation}
(\sigma v)^{\chi\chi}_{\rm{Majorana}} = v^2\,\frac{\lambda^4\,m_\chi^2\,(m_\chi^4 + m_\phi^4)}{48\pi\,(m_\chi^2 + m_\phi^2)^4} \,.
\label{eq14}
\end{equation}
Setting $\rm {m_{\chi}=1.5~TeV}$, we get the parameters that give the
correct relic density, which are also shown in Fig.~\ref{fig:md}.
The DM annihilation cross section is p-wave suppressed by the small
velocity of DM particles. The typical velocity of DM particles is about
$0.3 c$ at freeze-out and $\sim10^{-3} c$ at present. Therefore a boost
factor of $10^4 \sim 10^5$ is necessary. The Sommerfeld enhancement is
a natural mechanism that could enhance the DM annihilation cross section
for low relative velocities DM particles~\cite{Sommerfeld1931,
2005PhRvD..71f3528H,2009PhRvD..79a5014A}. Other mechanisms such as the
Breit-Wigner resonance was also proposed \cite{1991PhRvD..43.3191G,
1991NuPhB.360..145G}.

Then we consider the complex scalar DM model whose interaction can be
written as
\begin{equation}
{\cal L}_{\rm scalar} \supset \lambda_{i} X \overline{\psi^i}_L e_R^i
\, + \,h.c.,
\end{equation}
where $X$ denotes the DM particle and the mediator $\psi$ is a Dirac 
fermion with electric charge $-1$ and the corresponding lepton number. 
Here we only consider the complex scalar DM model because the direct 
detection rate can be effectively suppressed in this case 
\cite{2008PhRvD..78e6007B}. The cross section is
\begin{eqnarray}
\frac{1}{2}\,(\sigma v)^{\chi\chi^\dagger}_{\rm complex\, scalar} \,=\,
\frac{1}{2}\,\left[v^2\,\frac{\lambda^4\, m_X^2}{48\,\pi\,(m_X^2 +
m_\psi^2)^2} \right] \,.
\label{eq16}
\end{eqnarray}
The annihilation cross section is also p-wave suppressed. The coupling
and cross section correspond to the observed DM relic density are shown
in Fig.~\ref{fig:md}.

{\it Lepton Flavored DM model} ---
In the lepton portal DM model, DM particles annihilate into leptons
through lepton-flavored mediators. We can also assume that the DM particles 
have electron lepton number for Dirac and complex scalar DM model. 
In such a model, the DM particles could only annihilate into electrons 
and positrons. The Lagrangian is the same as the previous lepton portal 
DM models. To get the same annihilation cross section and hence the 
correct relic density with the previous model, the coupling parameter 
needs to be $\sqrt[4]{3}=1.3$ times larger according to Eqs. (\ref{eq12}) 
and (\ref{eq16}).

{\it TeV Right-handed Neutrino DM model} ---
One of the TeV right-handed neutrino models can be described by the
following Lagrangian \cite{2003PhRvD..67h5002K,2004PhRvD..69k3009C}
\begin{eqnarray}
{\cal L}_{\rm int} &=& f_{\alpha\beta} L^T_\alpha C i \tau_2 L_\beta S^+_1
   + g_{1\alpha} N_1 S_2^+ \ell_{\alpha R}
   + g_{2\alpha} N_2 S_2^+ \ell_{\alpha R}  + h.c. \nonumber \\
& + & \, M_{N_1} N^T_1 C N_1 + M_{N_2} N^T_2 C N_2 - V(S_1,S_2),
\label{lag}
\end{eqnarray}
where $N_{1,2}$ are right-handed neutrinos, $L_{\alpha,\beta}$ and
$l_{\alpha R}$ are the lepton doublet and singlet, $\alpha,\beta$ are
the family indices, $C$ is the charge-conjugation operator, $V(S_1,S_2)$
is the scalar potential which contains two complex scalar fields
($S_1$, $S_2$), and $f_{\alpha\beta}$ is the anti-symmetric coupling
matrix. The lighter one of the two right-handed neutrinos (denoted as
$N$ here) can be the DM candidate. In this model, the DM particles
annihilate into $e^+ e^-$, $\mu^+ \mu^-$ and $\tau^+ \tau^-$ through
$t$- and $u$-channel with an intermediate $S_2^+$. The cross section
is given by
\begin{eqnarray}
(\sigma v)_{\alpha} &=& \frac{g_{1\alpha}^4 }{64\pi}\, \frac{1}{s} \;
\int_{-1}^{1} \; d x \; \Biggr \{
 \frac{ s^2 ( 1- \beta_N x)^2}{ 4 \left[ M_N^2 - M_{S_2}^2 - \frac{s}{2}
 ( 1- \beta_N x) \right]^2 }  \nonumber \\
&+&\frac{ s^2 ( 1+  \beta_N x)^2}{ 4 \left[ M_N^2 - M_{S_2}^2 - \frac{s}{2}
 ( 1 + \beta_N x) \right]^2 } \\
&-& \frac{2 M_N^2 s}{\left[M_N^2 - M_{S_2}^2 - \frac{s}{2} ( 1- \beta_N x)
\right]
\left[M_N^2 - M_{S_2}^2 - \frac{s}{2} ( 1 + \beta_N x) \right] } \Biggr \}.
\nonumber
\end{eqnarray}
From the above equation we can see that $\sigma v \to 0$ for $v \to 0$,
since $\beta_N = \left( 1 -  4 M^2_N / s \right)^{1/2}\sim 0$.
The cross section is also $p$-wave suppressed, which is common for
identical Majorana fermions \cite{1983PhRvL..50.1419G}.

\section{Anisotropies}

\begin{figure*}[!htb]
\includegraphics[width=0.45\textwidth]{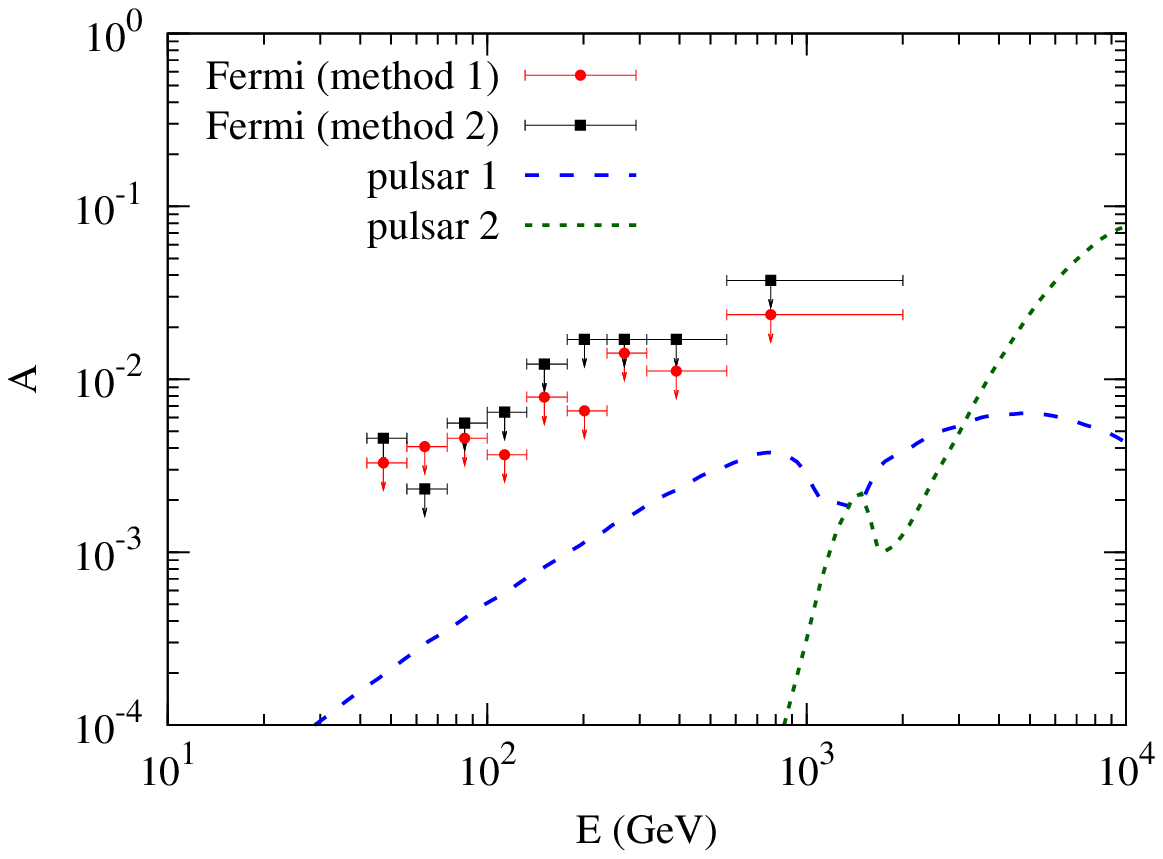}
\includegraphics[width=0.45\textwidth]{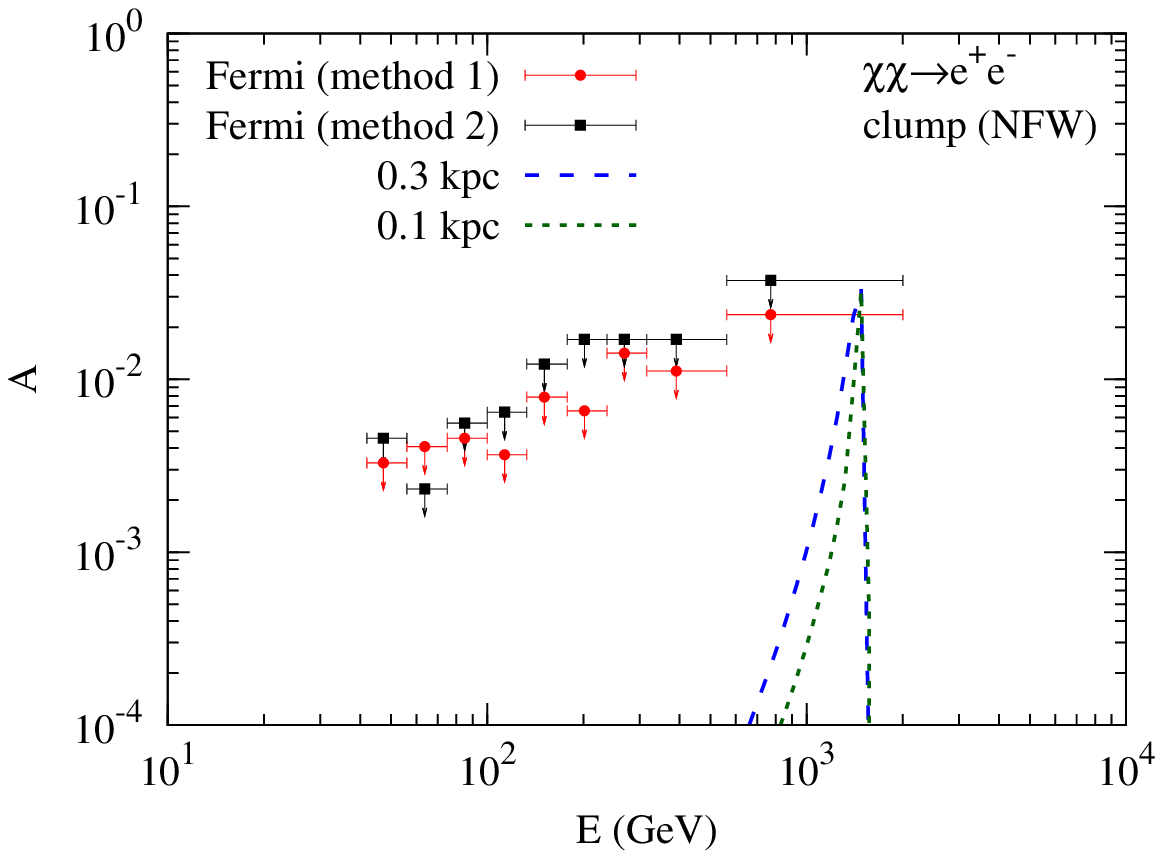}
\caption{Dipole anisotropy of $e^++e^-$ fluxes from the two-pulsar model
(left) as shown in Fig. \ref{fig:psr_real} and the DM subhalo model (right)
as shown in Fig. \ref{fig:elec_clump}. The 95\% upper limits from Fermi-LAT
observations are also shown \cite{2017PhRvL.118i1103A}.
\label{fig:ani}}
\end{figure*}

The anisotropies of high energy electrons can effectively test the
local origin models. The amplitude of the dipole anisotropy can be
calculated as
\begin{equation}
A=\frac{3D(E)}{c}\frac{|\nabla\phi_e|}{\phi_e},
\end{equation}
where $\phi_e$ is the local flux of $e^++e^-$. Assuming the background
is isotropic\footnote{The anisotropy amplitude of background $e^++e^-$
is about $6\times (10^{-4}-10^{-3})$ for energies between 10 GeV and
1 TeV \cite{2010PhRvD..82i2003A}.}, we calculate the anisotropies of
the pulsar model (see Fig. \ref{fig:psr_real}) and DM subhalo model
(see Fig. \ref{fig:elec_clump}). The results are shown in Fig. \ref{fig:ani}.

For the pulsar model, the anisotropies from both sources are about
$10^{-3}-10^{-2}$ at TeV energies. The high energy tails from pulsars
give relatively high anisotropies, although their fluxes are small.
This is because fresh particles injected just recently have smaller
diffusion distance, and hence give larger gradient than earlier injected
ones. We should keep in mind that the anisotropy of pulsar 1 is largely
uncertain, depending on the model parameters (mainly the age) assumed
\cite{2009APh....32..140G}. The age of pulsar 2 is better constrained,
and its anisotropy prediction is more robust. If the injection spectrum
of pulsar 2 is narrower than the Maxwellian distribution, then the age
of pulsar 2 can be smaller, and hence a larger anisotropy is possible.
For the DM subhalo model, the anisotropies can be as large as a few
percents around the DAMPE peak.

Compared with the upper limits obtained with seven years of Fermi-LAT
observations \cite{2017PhRvL.118i1103A}, the expected anisotropies for both 
models are consistent with the data. Future experiments such as the High 
Energy cosmic-Radiation Detection facility (HERD; \cite{2014SPIE.9144E..0XZ}) 
and the Cherenkov Telescope Array (CTA; \cite{2011ExA....32..193A}) may 
reach a sensitivity of $\sim10^{-3}$ around TeV energies and can effectively 
test these models \cite{2017arXiv170603745F,2013ApJ...772...18L}.

\section{Conclusion}

The precise measurements about the high energy electron (including positron) 
spectrum by DAMPE are very helpful in understanding the origin of CR 
electrons. In this work we extensively discuss the physical implications 
of the DAMPE data. Both the astrophysical models and the exotic DM 
annihilation/decay scenarios are examined. 
Our findings are summarized as follows.
\begin{itemize}
\item The spectral softening at $\sim0.9$ TeV suggests a cutoff (or break)
of the background electron spectrum, which is expected to be due to either
the discretness of CR source distributions in both space and time, or the
maximum energies of electron acceleration at the sources. The DAMPE data
enables a much improved determination of the cutoff energy of the background
electron spectrum, which is about $3$ TeV assuming an exponential form,
compared with the pre-DAMPE data.

\item Both the annihilation and decay scenarios of the simplified DM 
models to account for the sub-TeV electron/positron excesses are severely 
constrained by the CMB and/or $\gamma$-ray observations. Additional 
tuning of such models, through e.g., velocity-dependent annihilation, 
is required to reconcile with those constraints.

\item The tentative peak at $\sim1.4$ TeV suggested by DAMPE implies
that the sources should be close enough to the Earth ($\lesssim0.3$ kpc)
and inject nearly monochromatic electrons into the Galaxy. We find that the
cold and ultra-relativistic $e^+e^-$ wind from pulsars is a possible
source of such a structure. Our analysis further shows that the pulsar
should be middle-aged, relatively slowly-rotated, mildly magnetized,
and isolate in a density cavity (e.g., the local bubble).

\item An alternative explanation of the peak is the DM annihilation in
a nearby clump or a local density enhanced region. The distance of the
clump or size of the over-density region needs to be $\lesssim0.3$ kpc.
The required parameters of the DM clump or over-density are relatively
extreme compared with that of numerical simulations, if the annihilation
cross section is assumed to be $3\times10^{-26}$ cm$^3$ s$^{-1}$.
Specifically, a DM clump as massive as $10^7-10^8$ M$_{\odot}$ or a local
density enhancement of $17-35$ times of the canonical local density is
required to fit the data if the annihilation product is a pair of $e^+e^-$.
Moderate enhancement of the annihilation cross section would be helpful
to relax the tension between the model requirement and the N-body
simulations of the CDM structure formation. The DM clump model or local
density enhancement model is found to be consistent with the Fermi-LAT
$\gamma$-ray observations.

\item The expected anisotropies from either the pulsar model or the DM
clump model are consistent with the recent measurements by Fermi-LAT.
Future observations by e.g., CTA, will be able to detect such anisotropies
and test different models.

\end{itemize}

DAMPE will keep on operating for a few more years. More precise measurements
of the total $e^++e^-$ spectrum extending to higher energies are available
in the near future. Whether there are more structures in the high energy
window, which can critically distinguish the pulsar model from the DM one,
is particularly interesting. With more and more precise measurements, we
expect to significantly improve our understandings of the origin of CR
electrons.

\acknowledgments
This work is supported in part by the National Key Research and Development
Program of China (No. 2016YFA0400200), the National Natural Science Foundation 
of China (Nos. 11475189, 11525313, 11722328, 11773075), 
and the 100 Talents Program of Chinese Academy of Sciences. FL is also 
supported by the Youth Innovation Promotion Association of Chinese Academy 
of Sciences (No. 2016288).

\appendix*
\section{Density profiles of DM subhalos in the solar neighborhood}

We follow the results of high-resolution N-body simulation {\tt Aquarius}
to determine the density profile of a DM subhalo in the solar neighborhood
\cite{2008MNRAS.391.1685S}. The NFW profile was found to give reasonable
fit to the density profile of subhalos. The concentration of a subhalo,
defined as the mean overdensity within the radius at which the maximum
circular velocity is attained ($r_{\rm max}$) in units of the critical
density, is found to vary with both the subhalo mass and the spatial
location. For the solar neighborhood, we find approximately
\begin{equation}
\delta_V=1.2\times10^6\left(M_{\rm sub}/10^6\,M_{\odot}\right)^{-0.18},
\label{eq:con}
\end{equation}
where $M_{\rm sub}$ is the mass of a subhalo (after the tidal stripping).
The concentional NFW concentration parameter, $c\equiv r_v/r_s$ where
$r_v$ is the virial radius and $r_s$ is the scale radius, relates with
$\delta_V$ through
\begin{equation}
7.213~\delta_V=\delta_c=\frac{200}{3}\frac{c^3}{\ln(1+c)-c/(1+c)}.
\end{equation}

The tidal force from the main halo will remove the outer matter of a
subhalo, especially when it is close to the Milky Way center. This tidal
radius can be approximated as the radius at which the density is 0.02
times of the local average unbound density (i.e., $0.4$ GeV cm$^{-3}$
at the solar neighborhood). We find that the tidal radius is roughly
$0.2$ times of the original virial radius of a subhalo, and the enclose
mass is about half of the original mass of $M_{\rm sub}'$. So our
procedure to determine the density profile of a subhalo with tidal
stripping is as follows:
\begin{itemize}
\item Given a subhalo mass $M_{\rm sub}$, calculate its concentration
with Eq. (\ref{eq:con}).
\item Calculate the ``original'' density profile of the subhalo with
$M_{\rm sub}'=2M_{\rm sub}$ and the concentration.
\item Calculate the tidal radius $r_t$ of the subhalo, and remove the
DM outside of $r_t$.
\end{itemize}

\bibliographystyle{apsrev}
\bibliography{/home/yuanq/work/cygnus/tex/refs}

\end{document}